\documentclass[useAMS,usegraphicx,usenatbib]{mn2e} 
\usepackage{graphicx}
\usepackage{graphicx,float}
\usepackage{latexsym,amsmath,amssymb}
\usepackage{natbib}
\usepackage{times}
\usepackage{changebar}
\usepackage{subfigure}
\usepackage{ltxtable}
\usepackage{txfonts}
\usepackage{longtable, lscape}

\bibpunct{(}{)}{;}{a}{}{,}

\title[On the role of the $\Gamma-\lambda_{\rm\,Edd}$ relation on the X-ray Baldwin effect]{On the role of the $\Gamma-\lambda_{\rm\,Edd}$ relation on the X-ray Baldwin effect in Active Galactic Nuclei}
\author[C. Ricci et al.]{C. Ricci$^{1,2}$\thanks{E-mail:
ricci@kusastro.kyoto-u.ac.jp},  S. Paltani$^{1}$, Y. Ueda$^{2}$ and H. Awaki$^{3}$\\
$^{1}$Department of Astronomy, University of Geneva, ch. d'Ecogia 16, 1290 Versoix, Switzerland\\
$^{2}$Department of Astronomy, Kyoto University, Oiwake-cho, Sakyo-ku, Kyoto 606-8502\\
$^{3}$Department of Physics, Ehime University, Matsuyama, 790-8577, Japan }
\begin{document}
\date{Received; accepted}

\pagerange{\pageref{firstpage}--\pageref{lastpage}} \pubyear{2011}

\maketitle

\label{firstpage}

 
\begin{abstract}
The X-ray Baldwin effect is the inverse correlation between the equivalent width (EW) of the narrow component of the iron K$\alpha$ line and the X-ray luminosity of active galactic nuclei (AGN). A similar trend has also been observed between Fe K$\alpha$ EW and the Eddington ratio ($\lambda_{\mathrm{Edd}}$). Using {\it Chandra}/HEG results of Shu et al. (2010) and bolometric corrections we study the relation between EW and $\lambda_{\mathrm{Edd}}$, and find that $\log EW = (-0.13\pm0.03) \log \lambda_{\mathrm{Edd}} + 1.47 $. We explore the role of the known positive correlation between the photon index of the primary X-ray continuum $\Gamma$ and $\lambda _{\rm\,Edd}$ on the X-ray Baldwin effect. We simulate the iron K$\alpha$ line emitted by populations of unabsorbed AGN considering 3 different geometries of the reflecting material: toroidal, spherical-toroidal and slab. We find that the $\Gamma-\lambda_{\rm\,Edd}$ correlation cannot account for the whole X-ray Baldwin effect, unless a strong dependence of $\Gamma$ on $\lambda_{\rm\,Edd}$, such as the one recently found by \citet{Risaliti:2009nx} and \citet{Jin:2012fk}, is assumed. No clear correlation is found between EW and $\Gamma$. We conclude that a good understanding of the slope of the $\Gamma-\lambda_{\rm\,Edd}$ relation is critical to assess whether the trend plays a leading or rather a marginal role in the X-ray Baldwin effect.

\end{abstract}
               

   
\section{Introduction}

The first evidence of the existence of a relation between the luminosity and the equivalent width (EW) of a line in active galactic nuclei (AGN) was found by \citet{Baldwin:1977fk} for the broad C\,IV$\,\lambda1549$ emission line. This trend is usually called the {\it Baldwin effect}, and it has been found for several other optical, UV and IR lines (see \citealp{Shields:2007qf} for a review).
A possible explanation for the Baldwin effect of the lines produced in the broad line region (BLR) is that it is related to a luminosity-dependent spectral energy distribution (SED): if more luminous objects have a a softer UV/X-ray spectra, then ionisation and photoelectric heating of the gas in the BLR would be reduced (e.g., \citealp{Korista:1998dq}). An anti-correlation between the EW and the luminosity was found for the iron\,K$\alpha$ line in the X-ray band by \citeauthor{Iwasawa:1993ys} (\citeyear{Iwasawa:1993ys}, $EW\propto L^{-0.20}$), and it is usually dubbed the {\it X-ray Baldwin effect} or the {\it Iwasawa-Taniguchi effect}. Recent studies performed with the highest available spectral energy resolution of {\it Chandra}/HEG have shown that the X-ray Baldwin effect is mostly related to the narrow component of the Fe K$\alpha$ line, peaking at 6.4\,keV (e.g., \citealp{Shu:2010zr}). This component is thought to arise in the molecular torus, although a contribution from the BLR (e.g., \citealp{Bianchi:2008dz}) or from the outer part of the disk (e.g., \citealp{Petrucci:2002fk}) cannot be completely ruled out. The X-ray Baldwin effect has been often explained as being due to the decrease of the covering factor of the torus with the luminosity (e.g., \citealp{Page:2004kx}, \citealp{Bianchi:2007vn}), in the frame of the so-called luminosity-dependent unification models (e.g., \citealp{Ueda:2003qf}). In a recent paper \citep{Ricci:2013uq} we have shown that a luminosity-dependent covering factor of the torus is able to explain the slope of the X-ray Baldwin effect for equatorial column density of the torus $N_{\mathrm{H}}^{\mathrm{T}}\gtrsim 10^{23.1}\rm\,cm^{-2}$. Other possible explanations proposed are the delayed response of the reprocessing material with respect to the variability of the primary continuum \citep{Jiang:2006vn,Shu:2012fk}, or a change in the ionisation state of the iron-emitting material with the luminosity \citep{Nandra:1997fk,Nayakshin:2000uq}.

Studying a large sample of unabsorbed AGN, \citet{Bianchi:2007vn} found a highly significant anti-correlation also between the Fe K$\alpha$ EW and the Eddington ratio ($\lambda_{\mathrm{Edd}}=L_{\mathrm{Bol}}/L_{\mathrm{Edd}}$, where $L_{\mathrm{Edd}}=1.2\times10^{38}(M_{\rm\,BH}/M_{\sun})\rm\,erg\,s^{-1}$). \citet{Bianchi:2007vn} found $EW\propto \lambda_{\mathrm{Edd}}^{-0.19}$, i.e. an index similar to that obtained considering the X-ray luminosity, which leaves open the possibility that the Eddington ratio could play a significant role in the X-ray Baldwin effect. \citet{Winter:2009uq} found that the Fe K$\alpha$ EW is more strongly correlated to $\lambda_{\mathrm{Edd}}$ than to the luminosity. However, in their work they used both unabsorbed and absorbed AGN, which might add confusion to the relation, as the EW of significantly obscured ($N_{\rm\,H}> 10^{23}\rm\,cm^{-2}$) AGN is enhanced by the depletion of the continuum caused by the absorbing material in the line of sight.  In a recent study of {\it Chandra}/HEG unabsorbed AGN, \citet{Shu:2010zr}, using the ratio between the 2--10\,keV and the Eddington luminosity $L_{\,2-10}/L_{\mathrm{Edd}}$ as a proxy of $\lambda_{\mathrm{Edd}}$ (equivalent to assuming a constant bolometric correction), found that $EW\propto (L_{\,2-10}/L_{\mathrm{Edd}})^{-0.20}$. They also found that, similarly to what has been obtained with the luminosity, the correlation is weaker ($EW\propto (L_{\,2-10}/L_{\mathrm{Edd}})^{-0.11}$) when one considers values of $L_{\,2-10}/L_{\mathrm{Edd}}$ and EW averaged over different observations ({\it fit per source}) rather than using all the available observations for every object of the sample ({\it fit per observation}). In Table\,\ref{tab:XBEref} we report the results of the most recent studies of the $EW-\lambda_{\mathrm{Edd}}$ relation.

The first hint of the existence of an anti-correlation between the photon index $\Gamma$ and the full width half-maximum (FWHM) of $\mathrm{H}\beta$ was found in the 0.1--2.4\,keV energy band by \citet{Boller:1996kl} studying a sample of narrow-line Seyfert\,1s with {\it ROSAT}. This trend was later confirmed using {\it ASCA} observations at higher energies (e.g., \citealp{Brandt:1997fk}, \citealp{Reeves:2000uq}), and \citet{Brandt:1998kx} hypothesised that it might be related to the dependence of both $\mathrm{FWHM}(\mathrm{H}\beta)$ and $\Gamma$ on $\lambda_{\mathrm{Edd}}$. A significant positive correlation between $\Gamma$ and $\lambda_{\mathrm{Edd}}$ was later found by several works performed using {\it ASCA}, {\it ROSAT}, {\it Swift} and {\it XMM-Newton} (e.g., \citealp{Wang:2004ly}, \citealp{Grupe:2004fk}, \citealp{Porquet:2004vn}, \citealp{Bian:2005ve}, \citealp{Grupe:2010oq}). The degeneracy between the dependence of $\Gamma$ on $\mathrm{FWHM}(\mathrm{H}\beta)$ and on $\lambda_{\mathrm{Edd}}$ was broken by \citet{Shemmer:2006uq,Shemmer:2008fk}, who found that $\lambda_{\mathrm{Edd}}$ seems to be the main driver of the correlation. \citet{Shemmer:2008fk}, using estimates of $M_{\mathrm{BH}}$ obtained using H$\beta$, found that the relation between $\Gamma$ and $\lambda_{\mathrm{Edd}}$ can be approximated by

\begin{equation}\label{eq:lambda_gamma}
\Gamma\simeq0.31\log \lambda_{\mathrm{Edd}}+2.11,
\end{equation}
consistently with what was obtained by \citet{Wang:2004ly} and by \citet{Kelly:2007fk}. A steeper relation between $\Gamma$ and $\lambda_{\rm\,Edd}$ ($\Gamma \propto 0.60\,\log \lambda_{\rm\,Edd}$) has been found by \citet{Risaliti:2009nx} and \citet{Jin:2012fk}. The relation between $\Gamma$ and $\lambda_{\mathrm{Edd}}$ implies the existence of a strong link between the properties of the accretion flow and those of the warm corona. The fact that objects with higher values of $\lambda_{\mathrm{Edd}}$ show steeper photon indices has been explained by several authors as being due to a more efficient cooling of the corona caused by the larger optical/UV flux emitted by the accreting disk (e.g., \citealp{Shemmer:2008fk}). {This behaviour of $\Gamma$ is similar to what has been observed in stellar mass black holes \citep{Remillard:2006uq}, in agreement with the unification of accretion onto black holes across the mass scale.

The positive $\Gamma-\lambda_{\mathrm{Edd}}$ correlation might play a role on the X-ray Baldwin effect. Objects with lower values of $\lambda_{\mathrm{Edd}}$ have a flatter continuum, which implies that the amount of photons that can produce the fluorescent Fe K$\alpha$ emission is larger, which would produce a larger EW. The opposite would happen for objects at high Eddington ratios, which would create an anti-correlation between EW and $\lambda_{\mathrm{Edd}}$. If the EW of the iron K$\alpha$ line is mainly correlated to $\lambda_{\mathrm{Edd}}$, the relation between $\Gamma$ and $\lambda_{\mathrm{Edd}}$ could also be, at least in part, responsible for the observed trend. In this work we aim at constraining the effect of the $\Gamma-\lambda_{\mathrm{Edd}}$ correlation on the EW of the iron K$\alpha$ line for different geometries. We simulate the spectra of unabsorbed populations of AGN considering three different geometries for the reflecting material (spherical-toroidal, toroidal and slab), and we study the relation between the EW of the iron K$\alpha$ line and the Eddington ratio. We consider only unabsorbed AGN to compare our results to the most reliable studies of the X-ray Baldwin effect.

\begin{figure}
\begin{center}
\includegraphics[width=\columnwidth]{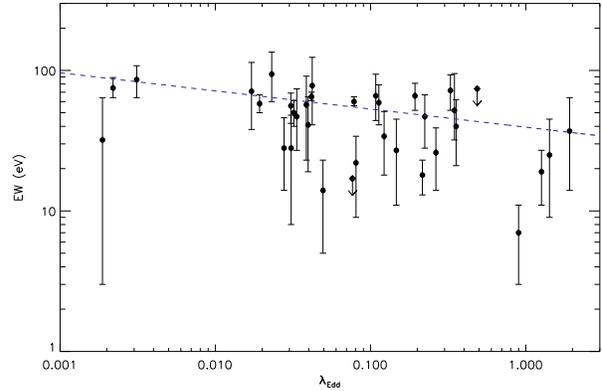}
 \caption{Equivalent width of the narrow component of the iron K$\alpha$ line versus the Eddington ratio for the objects of the {\it Chandra}/HEG sample of \citet{Shu:2010zr}. The values of the Eddington ratio were calculated using bolometric corrections obtained from the SED (Table\,\ref{tab:edd_ratio}). The dashed line represents the best fit to the data ($B=-0.13\pm0.03$, Table\,\ref{tab:XBEref}).}
\label{fig:ew_eddratio}
\end{center}
\end{figure}

\begin{table}
\caption[]{Summary of the most recent studies of the $EW-\lambda_{\mathrm{Edd}}$ relation. The model commonly used to fit the data is $\log EW=A+B \log \lambda_{\mathrm{Edd}}$, where $\lambda_{\mathrm{Edd}}=L_{\mathrm{Bol}}/L_{\mathrm{Edd}}$. The table lists separately works for which the fit was done {\it per observation} (i.e. using all the available observations for every source of the sample) and {\it per source} (i.e. averaging the values of EW and $\lambda_{\mathrm{Edd}}$ of different observations of the same source). The table shows (1) the reference, (2) the value of the slope, (3) the number and the type (radio-quiet, RQ, or radio-loud, RL) of objects of the sample, and (4) the observatory and instrument used.}
\label{tab:XBEref}
\begin{center}
\resizebox{\columnwidth}{!}{
\begin{tabular}{lccc}
\hline
\hline
\noalign{\smallskip}
\multicolumn{1}{c}{(1)} & (2) & (3) & (4) \\ 
\noalign{\smallskip}
Reference  & $B$ & Sample & Observatory/instrument\\ 
\noalign{\smallskip}
\hline
\noalign{\smallskip}
\multicolumn{4}{l}{\bf Fits per Observation} \\
\noalign{\smallskip}
\hline
\noalign{\smallskip}
\noalign{\smallskip}
\citealp{Shu:2010zr}$^{\,1}$        & $-0.20\pm0.03$ & 33 (RQ+RL) & {\it Chandra}/HEG \\  
\noalign{\smallskip}
  \citealp{Bianchi:2007vn}$^{\,2}$   & $-0.19\pm0.05$ & 82 (RQ) & {\it XMM-Newton}/EPIC \\                    		
\noalign{\smallskip}
  \citealp{Bianchi:2007vn}$^{\,1}$   & $-0.16\pm0.06$ & 82 (RQ) & {\it XMM-Newton}/EPIC \\                    		
\noalign{\smallskip}
\noalign{\smallskip}
\hline
\noalign{\smallskip}
\multicolumn{4}{l}{\bf Fits per Source} \\
\noalign{\smallskip}
\hline
\noalign{\smallskip}
\citealp{Shu:2010zr}$^{\,1}$        & $-0.11\pm0.04$ & 33 (RQ+RL) & {\it Chandra}/HEG \\                    
\noalign{\smallskip}
This work$^{\,3}$        & $-0.13\pm0.03$ & 33 (RQ+RL) & {\it Chandra}/HEG \\                    
\noalign{\smallskip}
This work$^{\rm\,3,\,A}$        & $-0.18\pm0.05$ & 33 (RQ+RL) & {\it Chandra}/HEG \\                    
\noalign{\smallskip}\hline
\noalign{\smallskip}
\multicolumn{4}{l}{{\footnotesize {\bf Notes}: $^1$ constant bolometric correction; $^2$ luminosity-dependent bolometric; }} \\
\multicolumn{4}{l}{{\footnotesize  corrections of \citet{Marconi:2004fk}; $^3$ using the average bolometric corrections }} \\
\multicolumn{4}{l}{{\footnotesize  calculated directly from the AGN bolometric emission (see Sect.\,\ref{sect:ewlambdaedd}); }} \\
\multicolumn{4}{l}{{\footnotesize  $^{\rm\,A}$ taking into account the errors on $\lambda_{\rm\,Edd}$. }} \\
\end{tabular}
}
\end{center}
\end{table}

\begin{figure*}\label{fig:geometries}
\centering
\begin{minipage}[!b]{.32\textwidth}
\centering
\includegraphics[angle=270,width=6cm]{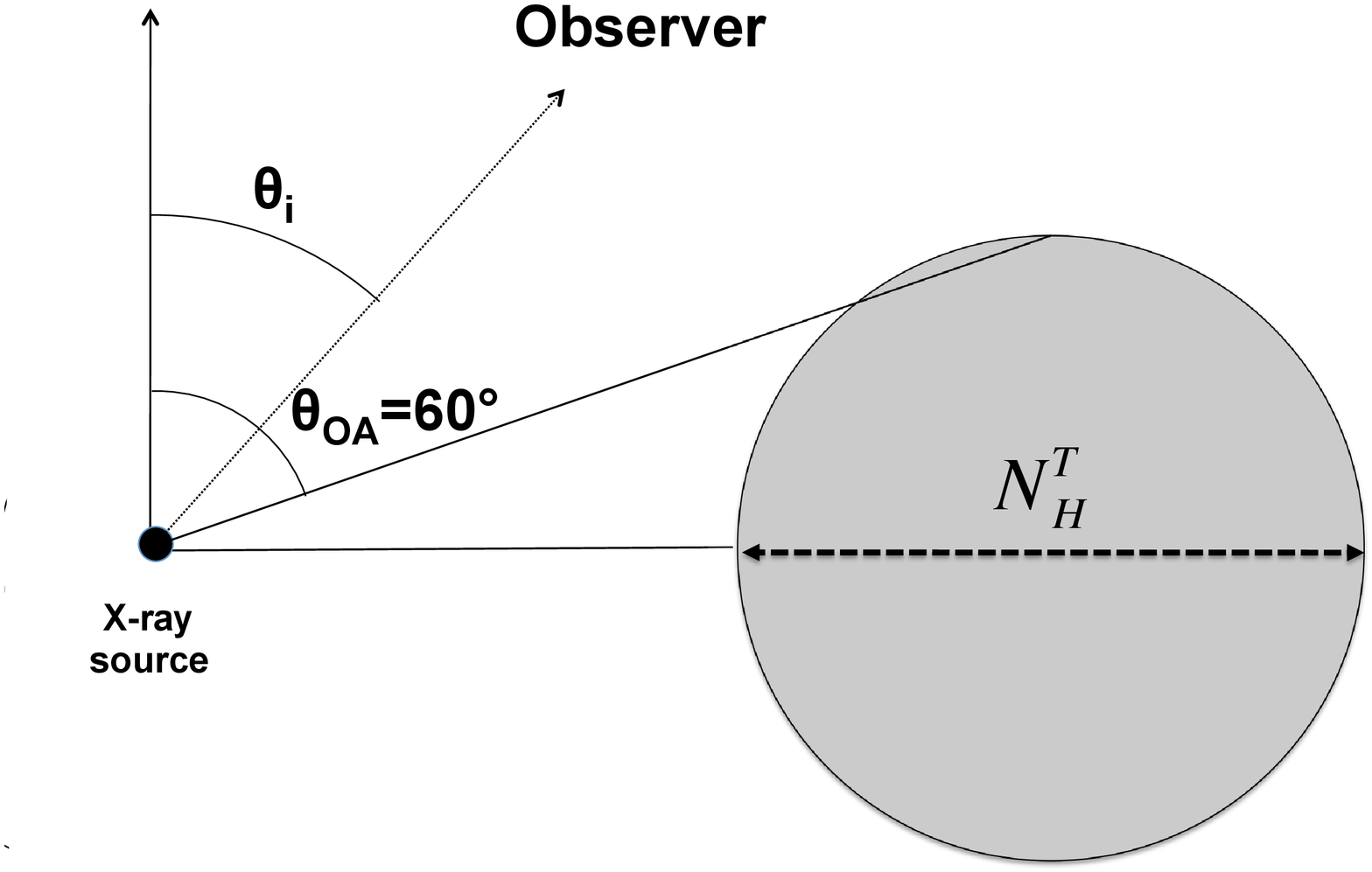}
\end{minipage}
\hspace{0.05cm}
\begin{minipage}[!b]{.32\textwidth}
\centering
\includegraphics[angle=270,width=6cm]{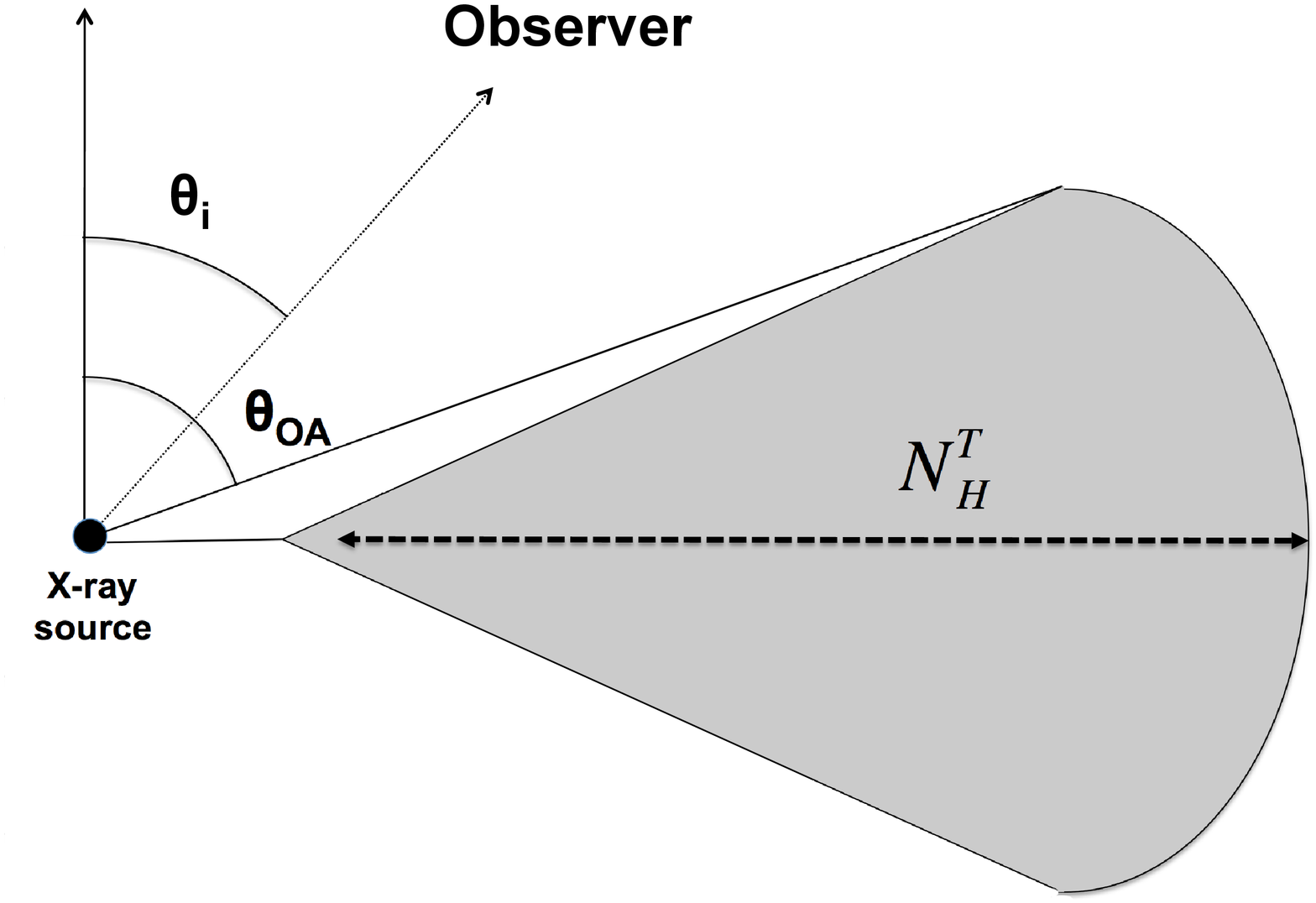}\end{minipage}
\hspace{0.05cm}
\begin{minipage}[!b]{.32\textwidth}
\centering
\includegraphics[angle=270,width=6cm]{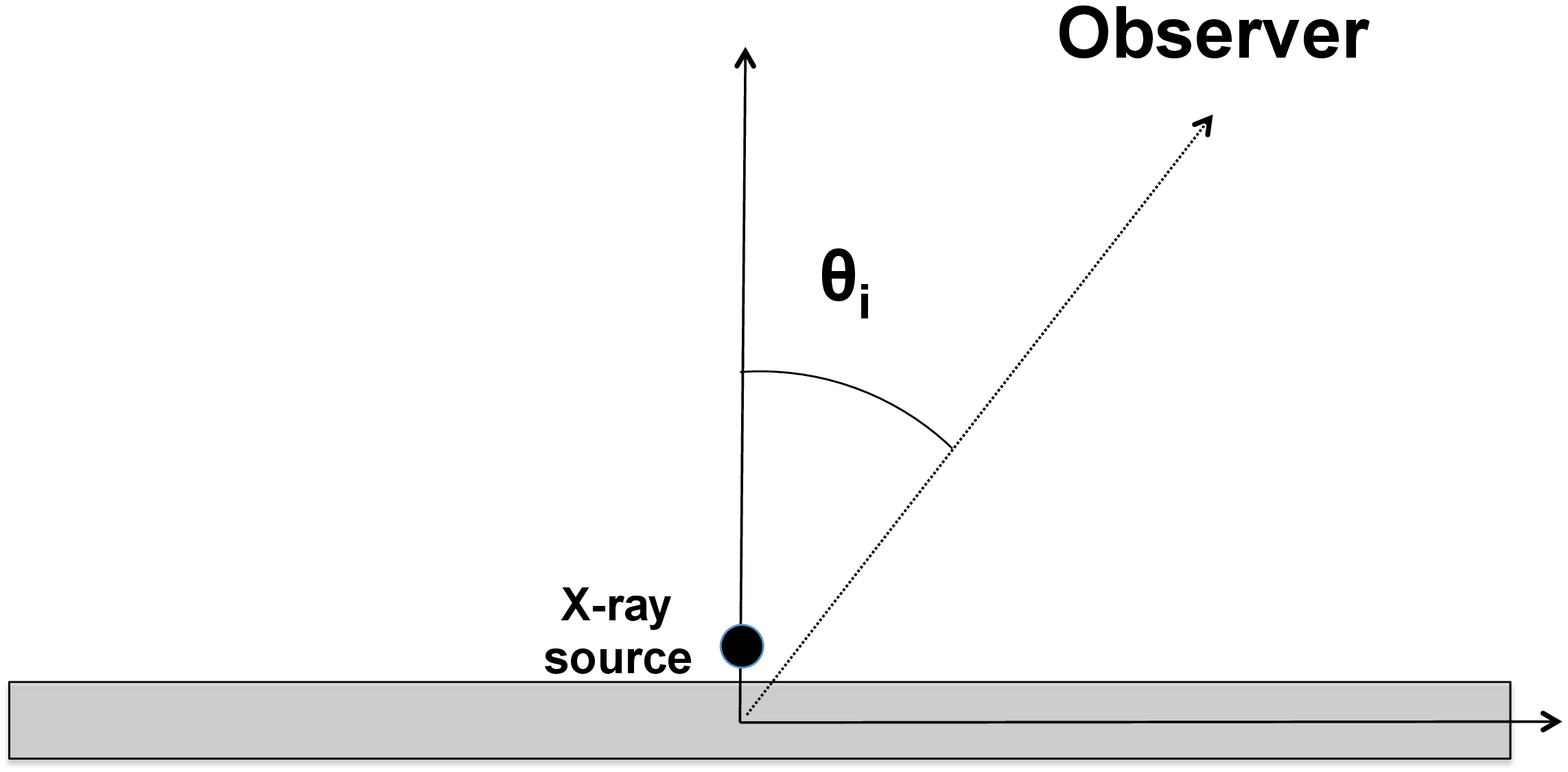}\end{minipage}
 \begin{minipage}[t]{1\textwidth}
  \caption{Schematic representation of the toroidal ({\it left panel}), spherical-toroidal ({\it central panel}) and slab ({\it right panel}) geometries. In all cases $\theta_{\rm\,i}$ is the inclination angle of the observer. In the toroidal geometry the half-opening angle $\theta_{\mathrm{OA}}$ is set to $60^{\circ}$, while in the spherical-toroidal geometry it varies between 10$^{\circ}$ and $70^{\circ}$. In both the toroidal and spherical-toroidal geometry the equatorial column density of the torus $N_{\rm\,H}^{\mathrm{\,T}}$ spans between $10^{22.5}$ and $10^{25}\rm\,cm^{-2}$.}
\label{fig:geometries}
 \end{minipage}

\end{figure*}

The paper is organised as follows. In Sect.\,\ref{sect:ewlambdaedd} we study the relation between EW and $\lambda_{\mathrm{Edd}}$ using the {\it Chandra}/HEG data of \citet{Shu:2010zr} and the bolometric corrections obtained from studies of the AGN spectral energy distribution. In Sect.\,\ref{sect:simulations} we present our simulations for a toroidal (\ref{sect:toroidal}), a spherical toroidal (\ref{sect:spherical_toroidal}) and a slab (\ref{sect:slab_geometry}) geometry, we discuss the influence of scatter (\ref{Sect:scatter}) and of the underlying $\lambda_{\mathrm{Edd}}$ distribution (\ref{Sect:edd_distr}). In Sect.\,\ref{Sect:EWvsGamma} we study the relation between the narrow Fe K$\alpha$ EW and the photon index using the {\it Chandra} sample of \citet{Shu:2010zr} and  the {\it Suzaku} sample of \citet{Fukazawa:2011fk}. In Sect.\,\ref{sect:discussion} we discuss our results, and in Sect.\,\ref{sect:conclusions} we present our conclusions.

\begin{table*}
\caption[]{The table reports (1) the equivalent width  of the narrow component of the iron K$\alpha$ line, (2) the 2--10\,keV luminosity, (3) the black hole mass, (4) the method used for calculating the black hole mass, (5) the 2--10\,keV average bolometric correction, (6) the bolometric luminosity, and (7) the Eddington ratio for the objects in the sample of \citet{Shu:2010zr}. The values of (1) and (2) were taken from \citet{Shu:2010zr} averaging when possible different observations. For the objects for which no value of the 2--10\,keV bolometric correction was available we fixed $\kappa_{\mathrm{x}}=20$.}
\label{tab:edd_ratio}
\begin{center}
\begin{tabular}{lccccccc}
\hline
\hline
\noalign{\smallskip}
& (1) & (2) & (3) & (4) & (5) & (6) & (7) \\ 
\noalign{\smallskip}
Source 						 & EW  &  $L_{\,2-10}$& $\log (M_{\mathrm{BH}}/M_{\sun})$ & Method  & $\kappa_{\mathrm{x}}$ & $\log L_{\mathrm{Bol}}$& $\lambda_{\rm\,Edd}$ \\ 
\noalign{\smallskip}
 & [{\scriptsize eV}]  &  [{\scriptsize $10^{43}\rm\,erg\,s^{-1}$}] &  &  &   & [{\scriptsize $\rm\,erg\,s^{-1}$}]&  \\ 

\noalign{\smallskip}
\hline
\noalign{\smallskip}
Fairall 9 	    					&$47_{-20}^{+27}$	 	&11.6	&$8.41 \pm 0.09^{\mathrm{a}}$	 &	RM &	$8.9 \pm   0.7$	&	   $45.0 $	 &	$    0.033 $   \\
\noalign{\smallskip}
NGC 526a	   			&$28_{-20}^{+20}$							&2.4		&$7.9 \pm 0.1^{\mathrm{b}}$	 & KB &	$13.1 \pm  0.9$	 &	   $44.5   $     &    $  0.031 $  \\
\noalign{\smallskip}
Mrk 590     					&$78_{-37}^{+46}$			&1.3		&$7.68 \pm 0.08^{\mathrm{a}} $  &	RM	 &$18.6 \pm  1.1$	 &	$44.4 $	 &	$    0.042 $   \\
\noalign{\smallskip}
NGC 985     					&$57_{-34}^{+34}$			&4.4		&$8.4 \pm 0.1^{\mathrm{b}}  	$ &	KB	 &$24.1\pm   1.3$	 &	$45.0  $	 &	$   0.039 $  \\
\noalign{\smallskip}
ESO 198$-$G24  			&$71_{-33}^{+43}$			&3.1		&$8.5 \pm 0.5^{\mathrm{c}} 	$ &	ER	 &$20^*$	 	&	$44.8 $	 &	$    0.017$   \\
\noalign{\smallskip}
3C 120    						&$47_{-19}^{+20}$		&11.7	&$7.7 \pm 0.2^{\mathrm{a}} $ &	RM	 &$12.6\pm   0.4$		&	$45.2 $	 &	$     0.22$    \\
\noalign{\smallskip}
NGC 2110    			      &$75_{-11}^{+14}$             &0.51        &$ 8.3 \pm 0.3^{\mathrm{d}} $ &	VD	 &$10.3\pm   0.7$	 &	$43.7 $	 &	$   0.0022$   \\ 
\noalign{\smallskip}
PG 0844$+$349  			&$52_{-20}^{+43}$	 	&5.4		&$8.0 \pm 0.2^{\mathrm{a}} $ &	RM	 &$72   \pm  8$			&	$45.6 $	 &	$     0.35$    \\
\noalign{\smallskip}
Mrk 705     					&$26_{-26}^{+48}$	 				&2.6		&$ 7.0 \pm 0.1^{\mathrm{e}} $ &	ER	 &$20^*$	 	&	$44.7 $	 &	$     0.49$   \\
\noalign{\smallskip}
MCG $-$5$-$23$-$16 			&$50_{-10}^{+11}$	 		&1.6		&$7.9\pm  0.4^{\mathrm{c}} $ &	VD	 &$20^*$	 		&	$44.5 $	 &	$    0.032$   \\
\noalign{\smallskip}
NGC 3227    					&$32_{-29}^{+32}$	 			&0.08	&$7.6\pm  0.2^{\mathrm{a}} $ &RM		 &$12.0   \pm  0.5$	 &	$43.0 $	 &	$   0.0019$   \\
\noalign{\smallskip}
NGC 3516    				&$58_{-8}^{+9}$	 	&0.70	&$7.6\pm  0.1^{\mathrm{a}} $ &	RM	 &$14.1 \pm   1.6	$	 &	$44.0 $	 &	$    0.019$  \\
\noalign{\smallskip}
NGC 3783    				&$60_{-4}^{+5}$	 	&1.4		&$7.47\pm  0.09^{\mathrm{a}} $ &	RM	 &$19.8\pm   0.9$	 &	$44.4 $	 &	$    0.078$   \\
\noalign{\smallskip}
NGC 4051   					&$94_{-34}^{+41}$	 	&0.02	&$6.3\pm  0.2^{\mathrm{a}} $ &	RM	 &$26.4\pm   1.1$	 &	$42.7 $	 &	$    0.023$    \\
\noalign{\smallskip}
NGC 4151   				&$65_{-4}^{+5}$	 	&0.36	&$7.1\pm  0.1^{\mathrm{a}} $ & RM		 &$18\pm   2$	 &	$43.8 $	 &	$    0.042$  \\
\noalign{\smallskip}
Mrk 766     					&$37_{-23}^{+27}$	 			&0.86	&$6.3\pm  0.4^{\mathrm{c}}	$ &	RM	 &$45\pm   3$	 	&	$44.6 $	 &	$    1.92$   \\
\noalign{\smallskip}
3C 273      				&$7_{-4}^{+4}$	 	&654	&$8.95\pm  0.07^{\mathrm{a}}	$ &	RM	 	&	$15\pm   3$	 &$47.0 $	 &	$     0.90$   \\
\noalign{\smallskip}
NGC 4593    					&$59_{-18}^{+20}$	 	&0.81	&$6.7\pm  0.6^{\mathrm{a}}	$ &	RM	 &$9.0     \pm 0.5$	 	&	$43.9 $	 &	$     0.11$   \\
\noalign{\smallskip}
MCG $-$6$-$30$-$15 		&$18_{-5}^{+5}$   &0.59&$6.7\pm  0.2^{\mathrm{d}}$ & 	  VD      &$22.0    \pm  1.0$&$44.1 $&$     0.22$   \\
\noalign{\smallskip}
IRAS 13349$+$2438		&$40_{-19}^{+22}$	 	&10.7	&$7.7\pm  0.5^{\mathrm{c}}$ &	ER	 &$20^*$	 	&	$45.3 $	 &	$     0.36$   \\
\noalign{\smallskip}
IC 4329A   					&$19_{-8}^{+8}$	 	&10.1	&$7.0\pm  0.8^{\mathrm{a}} 	$ &	RM	 &$15.0    \pm 1.0$	 &	$45.2 $	 &	$      1.26$  \\
\noalign{\smallskip}
Mrk 279     					&$66_{-22}^{+28}$	 	&2.9		&$7.5\pm  0.1^{\mathrm{a}}	$ &	RM	 &$15.5\pm  0.3 $	 &	$44.7 $	 &	$     0.11$   \\
\noalign{\smallskip}
NGC 5506    					&$66_{-14}^{+15}$	 	&0.55	&$6.6\pm  0.5^{\mathrm{f}} $ & VD		 &$16.8\pm  1.2 $	 &	$44.0 $	 &	$     0.19$   \\
\noalign{\smallskip}
NGC 5548     				&$56_{-14}^{+13}$ 	&1.9		&$7.83\pm  0.03^{\mathrm{a}} $ &	RM	 &$13.1\pm  1.5$	 &	$44.4 $	 &	$    0.031$   \\
\noalign{\smallskip}
Mrk 290     				&$27_{-16}^{+18}$	 	&3.2		&$7.39\pm  0.07^{\mathrm{g}} $ &	RM	 &$13.5\pm  0.9 $	 &	$44.6  $	 &	$     0.15$   \\
\noalign{\smallskip}
PDS 456     					&$4_{-4}^{+13}$	 				&37.3	&$8.9\pm  0.5^{\mathrm{h}} $ &	ER	 &$20^* $	 	&	$45.9 $	 &	$    0.08$ \\
\noalign{\smallskip}
E1821$+$643      				&$26_{-12}^{+13}$ 	&362.5	&$9.4\pm  0.5^{\mathrm{h}}  	$ &	ER	 &$20^* $	 	&	$46.9 $	 &	$     0.26$    \\
\noalign{\smallskip}
3C 382     				&$14_{-9}^{+9}$	 	&40.6	&$8.8 \pm 0.5^{\mathrm{h}}	$ &	ER	 &$10.1 \pm  0.6 $	 &	$45.6 $	 &	$    0.049$   \\
\noalign{\smallskip}
IRAS 18325$-$5926		&$5_{-5}^{+9}$		 			&2.3		&$ -    $	 	 &	--	 &$20^* $	 	&	$44.7 $ &	$  -$   \\
\noalign{\smallskip}
4C $+$74.26    			&$28_{-14}^{+18}$	 		&66.2	&$9.6 \pm  0.5^{\mathrm{d}}$ &	ER	 &$20^* $	 	&	$46.1 $ &	$    0.028$   \\
\noalign{\smallskip}
Mrk 509     					&$34_{-16}^{+17}$	 	&15.6	&$8.16 \pm0.03^{\mathrm{a}}	$ &	RM	 &$13.6\pm  0.6$	 	&	$45.3  $ &	$    0.12$  \\
\noalign{\smallskip}
NGC 7213    			      &$86_{-22}^{+22}$          &0.18        &$ 8.0 \pm0.3^{\mathrm{i}}  	$ &	VD	 &$20.3\pm  1.6$	 &	$43.6 $	 &	$   0.0031$   \\
\noalign{\smallskip}
NGC 7314    				&$41_{-22}^{+24}$	 		&0.17	&$6.1\pm 0.1 ^{\mathrm{b}}	$ &	KB	 &$3.9\pm  0.4$	 &	$42.8 $	 &	$    0.040$  \\
\noalign{\smallskip}
Ark 564     					&$25_{-16}^{+20}$ 			&3.7		&$6.3\pm 0.5^{\mathrm{c}} 	$ &	ER	 &$9.2  \pm 3.7 $	 	&	$44.5  $	 &	$     1.42$  \\
\noalign{\smallskip}
MR 2251$-$178     				&$22_{-13}^{+12}$ 			&25.3	&$8.8^{\mathrm{j}} 	$ &	KB	 &$22.0\pm  0.3 $	 &	$45.8 $	 &	$     0.081$  \\
\noalign{\smallskip}
NGC 7469     				&$72_{-20}^{+21}$	 	&1.5		&$7.09\pm 0.05^{\mathrm{a}}	$ &	RM	 &$32.2\pm  1.3$	 	&	$44.7 $	 &	$     0.33$    \\
\noalign{\smallskip}
\hline
\noalign{\smallskip}
\multicolumn{8}{l}{{\bf Notes.} The values of the $M_{\mathrm{BH}}$ were taken from: $^{\mathrm{a}}$ \citet{Peterson:2004uq}, $^{\mathrm{b}}$ \citet{Vasudevan:2010kx}, $^{\mathrm{c}}$ \citet{De-Marco:2013fk}, }\\
\multicolumn{8}{l}{ $^{\mathrm{d}}$ \citet{Beckmann:2009fk}, $^{\mathrm{e}}$ \citet{Vestergaard:2002vn}, $^{\mathrm{f}}$ \citet{Middleton:2008ys}, $^{\mathrm{g}}$ \citet{Denney:2010kx}, $^{\mathrm{h}}$ \citet{Wang:2009ly},  }\\
\multicolumn{8}{l}{$^{\mathrm{i}}$ \citet{Woo:2002ve}, $^{\mathrm{j}}$ \citet{Winter:2009uq}. The methods used for the mass calculation are: reverberation mapping (RM), }\\
\multicolumn{8}{l}{ K-band host bulge luminosity (KB), empirical $L(5100\AA)$ vs $R_{\mathrm{BLR}}$ relation (ER), and stellar velocity dispersion (VD). }
\end{tabular}        
\end{center}
\end{table*}

\section{The $EW-\lambda_{\mathrm{Edd}}$ relation}\label{sect:ewlambdaedd}
The studies of the $EW-\lambda_{\mathrm{Edd}}$ relation performed so far have used either constant (e.g., \citealp{Shu:2010zr}, \citealp{Bianchi:2007vn}) or luminosity-dependent (e.g., \citealp{Marconi:2004fk}, \citealp{Bianchi:2007vn}) 2--10\,keV bolometric corrections ($\kappa_{\mathrm{x}}$, where $L_{\mathrm{Bol}}=\kappa_{\mathrm{x}}\cdot L_{\,2-10}$). It has been however shown that the value of $\kappa_{\mathrm{x}}$ does not have a clear dependence on the bolometric luminosity (e.g., \citealp{Vasudevan:2007fk}, \citealp{Marchese:2012fk}), while it appears to be connected to the Eddington ratio \citep{Vasudevan:2007fk}.

In order to better constrain the slope of the $EW-\lambda_{\mathrm{Edd}}$ we used the {\it per source} values of EW and $L_{\,2-10}$ reported by the {\it Chandra}/HEG study of \citet{Shu:2010zr} (where the line width was fixed to $\sigma=1$\,eV to consider only the narrow component), together with the best estimates of the black-hole masses of their samples reported in the literature. For 18 sources values of $M_{\mathrm{BH}}$ obtained by reverberation mapping were available (e.g., \citealp{Peterson:2004uq}), while for the rest of the sources we used values obtained by i) the K-band host bulge luminosity-$M_{\mathrm{BH}}$ relation (e.g., \citealp{Vasudevan:2010kx}), ii) the empirical $L(5100\AA)$ vs $R_{\mathrm{BLR}}$ relation (e.g., \citealp{Vestergaard:2002vn}), iii) the stellar velocity dispersion (e.g., \citealp{Woo:2002ve}). We took the values of $\kappa_{\mathrm{x}}$ obtained directly from studies of the bolometric AGN emission performed by \citet{Vasudevan:2007fk}, \citet{Vasudevan:2009uq}, \citet{Vasudevan:2009vn}, and \citet{Vasudevan:2010kx}, averaging the values when several were available. For the eight sources of the sample for which no value of $\kappa_{\mathrm{x}}$ was available we fixed $\kappa_{\mathrm{x}}=20$ \citep{Vasudevan:2009vn}. The values of $\lambda_{\mathrm{Edd}}$ for the {\it Chandra}/HEG sample are reported in Table\,\ref{tab:edd_ratio}. We fitted the data using
\begin{equation}\label{Eq:fit}
\log EW=A+B \log \lambda_{\mathrm{Edd}}.
\end{equation}
The scatter plot of EW versus $\lambda_{\mathrm{Edd}}$ and the fit to the data are shown in Fig.\,\ref{fig:ew_eddratio}. We found that using the measured bolometric correction the slope of $EW-\lambda_{\mathrm{Edd}}$ is consistent ($B=-0.13\pm0.03$) with the value obtained using constant values of $\kappa_{\mathrm{x}}$ (see Table\,\ref{tab:XBEref}). Using the \texttt{fitexy} procedure \citep{Press:1992fk}, which allows to take into account also the uncertainties of $\lambda_{\rm\,Edd}$, we obtained a steeper slope ($B=-0.18\pm0.05$). The correlation yields a null-hypothesis probability of $P_{\mathrm{n}}\sim 1\%$ and the Spearman's rank correlation coefficient is $\rho=-0.44$. Interestingly, the decrease of the EW with the X-ray luminosity is more significant ($P_{\mathrm{n}}\sim 0.03\%$, $\rho=-0.64$). Considering only the objects for which it was possible to estimate $\kappa_{\mathrm{x}}$ does not increase the significance of the correlation.

\section{Simulations}\label{sect:simulations}  
In order to simulate the $EW-\lambda_{\mathrm{Edd}}$ relation produced by the dependence of $\Gamma$ on $\lambda_{\mathrm{Edd}}$, we followed what we have done in \citet{Ricci:2013uq} to study the influence of the decrease of the covering factor of the torus with the luminosity on the X-ray Baldwin effect. We initially considered an uniform distribution of Eddington ratios, with $\lambda_{\mathrm{Edd}}$ spanning between $0.01$ and $1$, with a binning of $\Delta \lambda_{\mathrm{Edd}}=0.01$. For each value of $\lambda_{\mathrm{Edd}}$ we determined the photon index using the relation of \citeauthor{Shemmer:2008fk} (\citeyear{Shemmer:2008fk}, Eq.\,\ref{eq:lambda_gamma}), and simulated in XSPEC 12.7.1 \citep{Arnaud:1996kx} a large number of spectra, using Monte Carlo simulations of reflected X-ray radiation. These simulations were carried out to examine the shape of the reflected emission for a given X-ray continuum, and assume three different geometries of the reprocessing material (Fig.\,\ref{fig:geometries}): toroidal (Sect.\,\ref{sect:toroidal}), spherical-toroidal (Sect.\,\ref{sect:spherical_toroidal}) and slab (Sect.\,\ref{sect:slab_geometry}). For each geometry we explored the parameter space and created synthetic unabsorbed populations. In XSPEC we measured the value EW of the simulated spectra, using for the continuum the same models as those used for the simulations and a Gaussian line for the iron K$\alpha$ line. We then studied the relation of EW with the Eddington ratio fitting the values of the synthetic populations with Eq.\,\ref{Eq:fit} using the weighted least-square method. The weights used are $w=\sin \theta_{\rm\,i}$, where $\theta_{\rm\,i}$ is the inclination angle of the observer. This is done to take into account the non-uniform probability of randomly observing the AGN within a certain solid angle from the polar axis. In Sect.\,\ref{Sect:scatter} we study the influence of scatter in the $\Gamma-\lambda_{\mathrm{Edd}}$ relation, and in Sect.\,\ref {Sect:edd_distr} we assume more realistic $\lambda_{\mathrm{Edd}}$ distributions. In Sect.\,\ref{sect:discussion} we discuss the impact of a steeper $\Gamma-\lambda_{\mathrm{Edd}}$ trend, as that found by \citet{Risaliti:2009nx}, on our simulations. Throughout the paper we assume solar metallicities.

\begin{figure*}
\centering
\begin{minipage}[!b]{.48\textwidth}
\centering
\includegraphics[width=9cm]{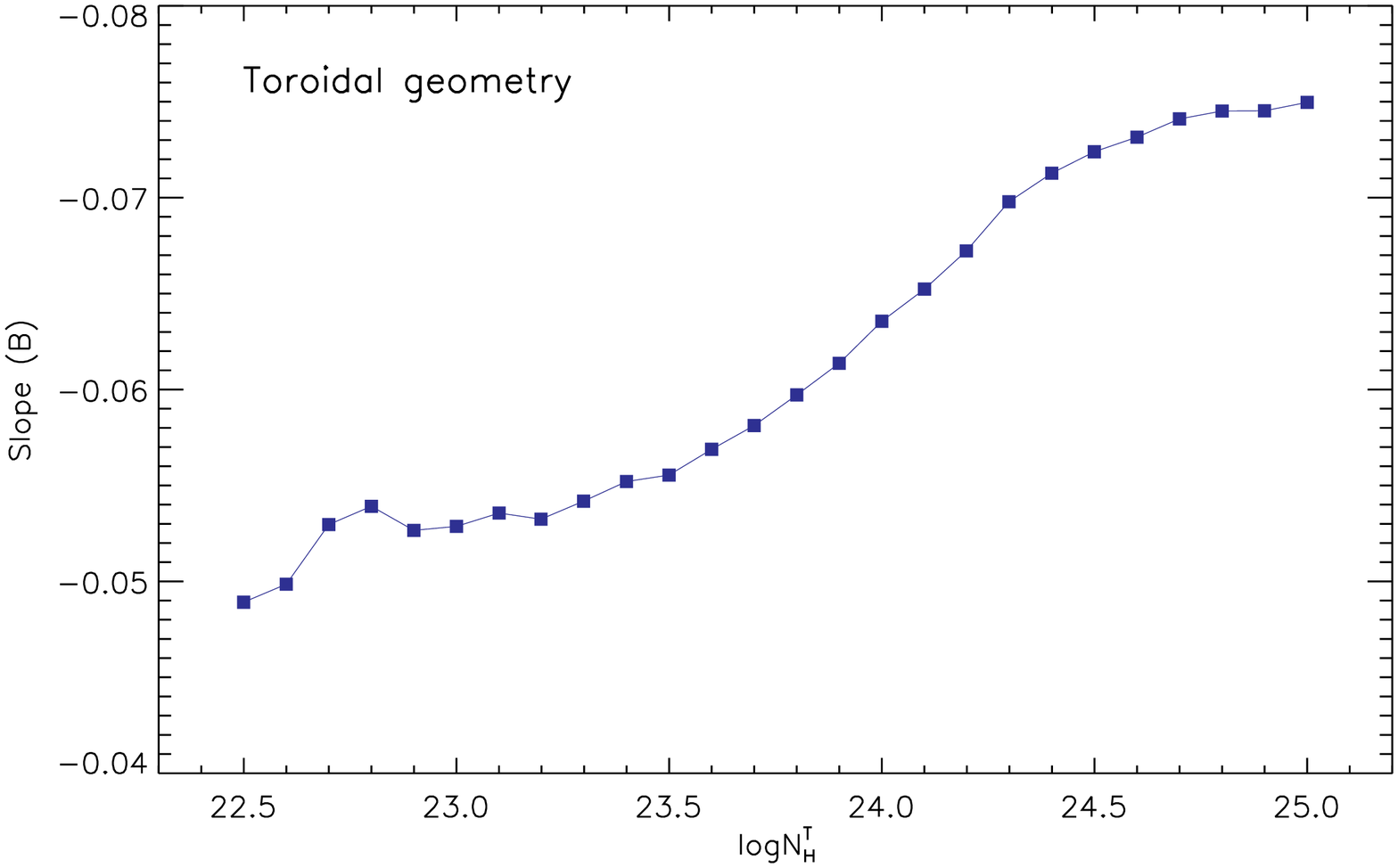}
\end{minipage}
\hspace{0.05cm}
\begin{minipage}[!b]{.48\textwidth}
\centering
\includegraphics[width=9cm]{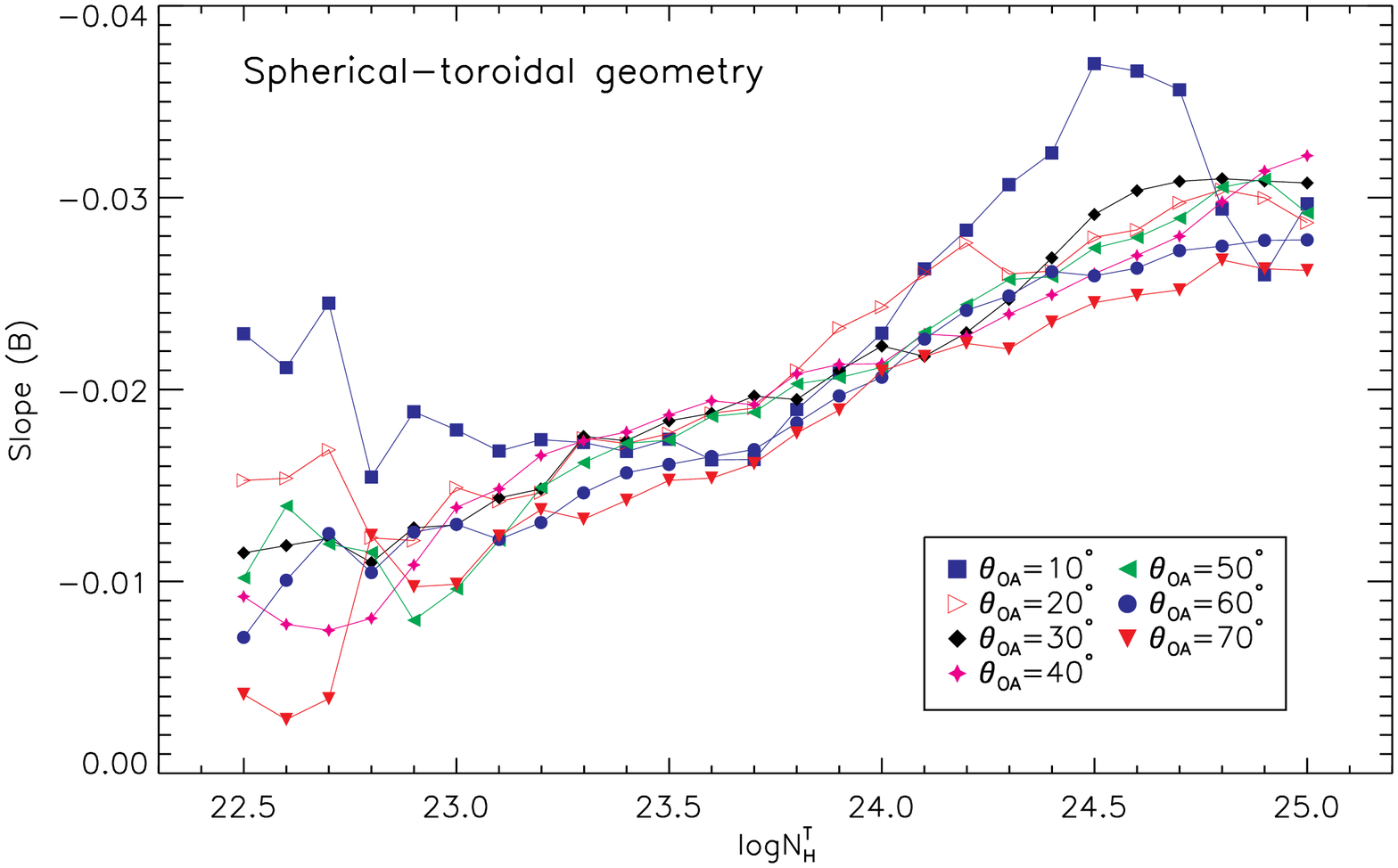}\end{minipage}
 \begin{minipage}[t]{1\textwidth}
  \caption{Slopes of $EW-\lambda_{\rm\,Edd}$ trend obtained simulating a population of unabsorbed AGN with photon indices increasing with the Eddington ratio as predicted by Eq.\,\ref{eq:lambda_gamma} for equatorial column densities of the torus between $10^{22.5}\rm\,cm^{-2}$ and $10^{25}\rm\,cm^{-2}$. The geometry considered here are toroidal ({\it left panel}) and spherical-toroidal ({\it right panel}).}
\label{fig:slopes}
 \end{minipage}

\end{figure*}

\subsection{Toroidal geometry} \label{sect:toroidal}
\citet{Murphy:2009uq} carried out Monte-Carlo calculations of Green's functions to study the reprocessed features originated in a toroidal geometry (see left panel of Fig.\,\ref{fig:geometries}). These simulations are part of a X-ray spectral fitting model (\texttt{MYTorus}\footnote{http://www.mytorus.com/}), which can be implemented in XSPEC as a combination of three table models. These table models include the zeroth-order continuum, the scattered continuum and a component which contains the fluorescent lines. The main parameters of \texttt{MYTorus} are the photon index $\Gamma$, the equatorial column density of the torus $N_{\rm\,H}^{\mathrm{\,T}}$ and the inclination angle of the observer $\theta_{\rm\,i}$. In \texttt{MYTorus} the half-opening angle of the torus $\theta_{\mathrm{OA}}$ is fixed to $60^{\circ}$, while $\theta_{\rm\,i}$ can vary between $0^{\circ}$ and $90^{\circ}$, and $N_{\rm\,H}^{\mathrm{\,T}}$ between $10^{22}\rm\,cm^{-2}$ and $10^{25}\rm\,cm^{-2}$.
The fluorescent emission lines in \texttt{MYTorus} are convolved with a Gaussian function (\texttt{gsmooth} in XSPEC) to take into account the velocity broadening. The width of the Gaussian broadening is
\begin{equation}\label{eq:gsmooth1}
\sigma_{\mathrm{E}}=\sigma_{\,\mathrm{L}}\left(\frac{E}{6\rm\,keV}\right)^{\alpha},
\end{equation}
where
\begin{equation}\label{eq:gsmooth2}
\sigma_{\,\mathrm{L}}=0.850\left(\frac{V_{\mathrm{FWHM}}}{100\rm\,km\,s^{-1}}\right)\rm\,eV.
\end{equation}
We used $\mathrm{FWHM}=2000\rm\,km\,s^{-1}$, as obtained by recent {\it Chandra}/HEG observations of unobscured AGN \citep{Shu:2010zr}, which is equivalent to $\sigma_{\,\mathrm{L}}=17\rm\,eV$, and $\alpha=1$. 
\texttt{MYTorus} does not allow a cut-off power law, but considers a {\it termination energy} ($E_{\mathrm{T}}$). As the choice of $E_{\mathrm{T}}$ does not affect significantly the values of the EW of the iron K$\alpha$ line, we arbitrarily chose $E_{\mathrm{T}}=500\rm\,keV$. We used a model which includes the three components implemented in \texttt{MYTorus}\footnote{In XSPEC the model was \texttt{pow*etable\{mytorus\_Ezero\_v00.fits\}
+atable\{mytorus\_scatteredH500\_v00.fits\}
+gsmooth(atable\{mytl\_V000010nEp000H500\_v00.fits\})}}, and fixed the value of the normalisation of the scattered continuum (which includes the Compton hump) and of the fluorescent-line model to that of the zeroth-order continuum.  We considered 26 values of equatorial column densities between $\log N_{\rm\,H}^{\mathrm{\,T}}=22.5$ and $\log N_{\rm\,H}^{\mathrm{\,T}}=25$, with a binning of $\Delta \log N_{\rm\,H}^{\mathrm{\,T}}=0.1$. For each value of $N_{\rm\,H}^{\mathrm{\,T}}$ we simulated a population of unabsorbed AGN, with $0^{\circ} \leq \theta_{\rm\,i} < \theta_{\mathrm{OA}}$, with a binning of $\Delta \theta_{\rm\,i}=3^{\circ}$, and with $\Gamma$ following Eq.\,\ref{eq:lambda_gamma}.

After extracting the values of EW we fitted the obtained $EW(\lambda_{\mathrm{Edd}})$ trend with Eq.\,\ref{eq:lambda_gamma} for each synthetic population, and obtained values of the slope that vary between $B\simeq -0.05$ (for $\log N_{\rm\,H}^{\mathrm{\,T}}=22.5$) and $B\simeq -0.075$ (for $\log N_{\rm\,H}^{\mathrm{\,T}}=25$). The variation of $B$ with $N_{\rm\,H}^{\mathrm{\,T}}$ is shown in the left panel of Fig.\,\ref{fig:slopes}.

\subsection{Spherical-toroidal geometry}\label{sect:spherical_toroidal}
Monte-Carlo simulations of the reprocessed radiation considering a spherical-toroidal geometry have been presented by \citet{Ikeda:2009nx}. These simulations are also stored in tables, so that they can be used for spectral fitting as in the case of \texttt{MYTorus}. The angles in this model are the same as those of the toroidal geometry (see central panel of Fig.\,\ref{fig:geometries} or Fig.\,1 in \citealp{Ricci:2013uq}). The model of \citet{Ikeda:2009nx} has the advantage, with respect to \texttt{MYTorus}, of having a variable half-opening angle of the torus $\theta_{\mathrm{OA}}$. The other parameters of this model are $\theta_{\rm\,i}$, $\Gamma$ and $N_{\rm\,H}^{\mathrm{\,T}}$, similarly to \texttt{MYTorus}, although the values of $\theta_{\rm\,i}$ allowed are in a slightly different range ($1^{\circ}-89^{\circ}$). In the model the energy of the cutoff is fixed to $E_C=360\rm\,keV$.
We simulated an unabsorbed population of AGN for each value of $\theta_{\mathrm{OA}}$ and $\log N_{\rm\,H}^{\mathrm{\,T}}$, using values of $\theta_{\rm\,i}$ between $1^{\circ}$ and $\theta_{\mathrm{OA}}$ with a binning of $\Delta \theta_{\rm\,i}=3^{\circ}$, and the photon index varying according to Eq.\,\ref{eq:lambda_gamma}. We selected 7 different values of $\theta_{\mathrm{OA}}$, from $10^{\circ}$ to $70^{\circ}$, with $\Delta \theta_{OA}=10^{\circ}$, and 26 values of $N_{\rm\,H}^{\mathrm{\,T}}$, from $\log N_{\rm\,H}^{\mathrm{\,T}}=22.5$ to $\log N_{\rm\,H}^{\mathrm{\,T}}=25$, with $\Delta \log N_{\rm\,H}^{\mathrm{\,T}}=0.1$. 

The values of $B$ obtained fitting with Eq.\,\ref{eq:lambda_gamma} the synthetic populations are shown in the right panel of Fig.\,\ref{fig:slopes}, and vary between $B\simeq-0.003$ (for $\theta_{\mathrm{OA}}=70^{\circ}$ and $\log N_{\rm\,H}^{\mathrm{\,T}}$=22.6) and $B\simeq-0.037$ (for $\theta_{\mathrm{OA}}=10^{\circ}$ and $\log N_{\rm\,H}^{\mathrm{\,T}}$=24.5).

\begin{figure*}
\centering
\includegraphics[width=\textwidth]{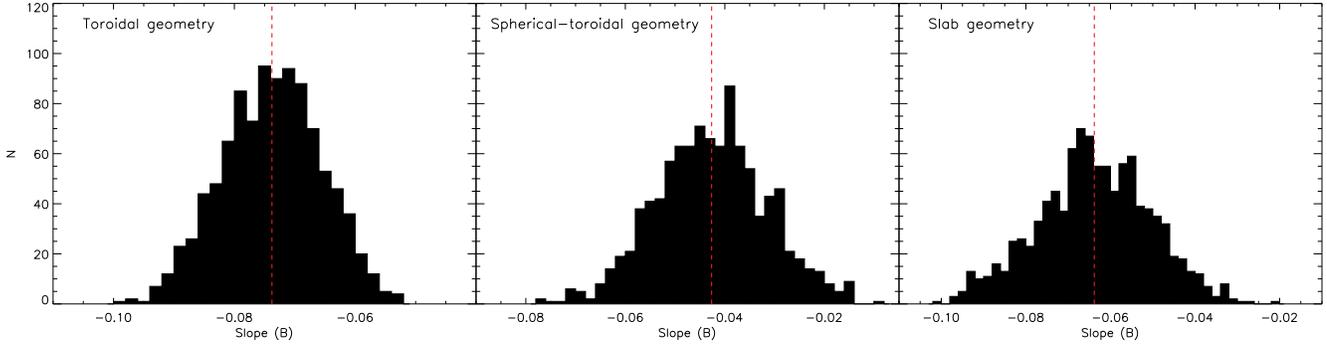}
  \caption{Histograms of the values of the slope of the $EW-\lambda_{\rm\,Edd}$ relation obtained by performing 1,000 Montecarlo simulations for the different geometries of the reflecting material described in Sects.\,\ref{sect:toroidal}-\ref{sect:slab_geometry}, including a scatter of $\Delta \Gamma\sim 0.1\,\Gamma$ in the $\Gamma-\lambda_{\rm\,Edd}$ relation (Eq.\,\ref{eq:lambda_gamma}) and considering random values of $\theta_{\rm\,i}$. For the toroidal geometry $\theta_{\mathrm{OA}}=60^{\circ}$, $N_{\rm\,H}^{\rm\,T}=10^{25}\rm\,cm^{-2}$ ({\it left panel}), for the spherical-toroidal geometry $\theta_{\mathrm{OA}}=10^{\circ}$, and $N_{\rm\,H}^{\rm\,T}=10^{24.5}\rm\,cm^{-2}$ ({\it central panel}), and for the slab geometry $R=1$ ({\it right panel}). The vertical dotted lines represent the average values of the slope. }
\label{fig:scatter}
\end{figure*}

\subsection{Slab geometry}\label{sect:slab_geometry}
For the sake of completeness we investigated also the impact of the $\Gamma$-$\lambda_{\mathrm{Edd}}$ relation on the Fe K$\alpha$ line EW using a slab geometry (right panel of Fig.\,\ref{fig:geometries}), commonly used to reproduce the reflection of the X-ray continuum on the accretion disk. We used the \texttt{pexmon} model \citep{Nandra:2006fk}, which is based on the spectral model for reflection on a semi-infinite slab of neutral material of \citeauthor{Magdziarz:1995fk} (\citeyear{Magdziarz:1995fk}, \texttt{pexrav}), linking to the reflection fraction $R$ the EW of the iron K$\alpha$ line. The parameter $R$ is defined as the strength of the reflection component relative to that expected from a slab subtending 2$\pi$ solid angle ($R= 2\pi/\Omega$). The strength of the Fe K$\alpha$ line in \texttt{pexmon} is determined using the Monte-Carlo calculations of \citet{George:1991fk}. In the model the relation between EW and $\Gamma$ is parametrised as
\begin{equation}\label{pexmon1}
EW=9.66\,EW_{0}\,(\Gamma^{-2.8}-0.56),
\end{equation}
where $EW_{0}=144\rm\,eV$ is the equivalent width for a face-on slab ($\theta_{\rm\,i}=0^{\circ}$), considering $\Gamma=1.9$ and iron abundance relative to hydrogen of $3.31\times 10^{-5}$ \citep{Anders:1982uq}. We simulated populations of unabsorbed AGN for five different values of the reflection parameter ($R=0.2,0.4,0.6,0.8,1$), using the same $\theta_{\rm\,i}$ distribution we have used for \texttt{MYTorus} ($\theta_{\mathrm{i}}< 60^{\circ}$). The energy of the cutoff does not play a significant role on the EW of the Fe K$\alpha$ line, and it was fixed to $E_C=500\rm\,keV$. In the slab model the value of $N_{\rm\,H}$ of the reflecting material is fixed to infinite, so that we cannot vary it.

Fitting the simulated data we obtain that the value of the slope does not vary significantly for different values of $R$, and is $B\simeq -0.058$.

\subsection{The effects of scatter in the $\Gamma-\lambda_{\mathrm{Edd}}$ relation}\label{Sect:scatter}
The values of the slopes obtained in the previous sections for different geometries of the reflector have been calculated neglecting the significant scatter found in the $\Gamma-\lambda_{\mathrm{Edd}}$ relation ($\Delta \Gamma\sim 0.1\,\Gamma$, \citealp{Shemmer:2008fk}). In order to account for this scatter, for each of the three geometries, and considering the set of parameters for which slope was the steepest, we carried out 1,000 Monte-Carlo simulations. In these simulations, for each value of $\lambda_{\mathrm{Edd}}$ the photon index had a random scatter within $\Delta \Gamma\sim 0.1\,\Gamma$. The inclination angle was randomly selected, in order to simulate unabsorbed AGN, between $1^{\circ}$ and $\theta_{\mathrm{OA}}$ for the spherical-toroidal geometry, and between $0^{\circ}$ and $60^{\circ}$ for the remaining two geometries. For each simulation we calculated the value of $B$ using Eq.\,\ref{Eq:fit} as described in Sect.\,\ref{sect:simulations}.

For the toroidal geometry ($\theta_{\mathrm{OA}}=60^{\circ}$, $N_{\rm\,H}^{\rm\,T}=10^{25}\rm\,cm^{-2}$) the average slope is $B=-0.074$ and the standard deviation $\sigma_{\rm\,B}=0.008$ (left panel of Fig.\,\ref{fig:scatter}), for the spherical-toroidal geometry ($\theta_{\mathrm{OA}}=10^{\circ}$, $N_{\rm\,H}^{\rm\,T}=10^{24.5}\rm\,cm^{-2}$) the average slope is $B=-0.043$ and the standard deviation $\sigma_{\rm\,B}=0.011$ (central panel of Fig.\,\ref{fig:scatter}), while for the slab geometry the average slope we obtained is $B=-0.064$ and the standard deviation $\sigma_{\rm\,B}=0.013$. None of the Monte-Carlo runs produce values of $B$ consistent with the slope of the X-ray Baldwin effect we found in Sect.\,\ref{sect:ewlambdaedd} (see also Table\,\ref{tab:XBEref}).

\subsection{Assuming a realistic $\lambda_{\mathrm{Edd}}$ distribution}\label{Sect:edd_distr}
The simulations discussed so far have been carried out assuming an unrealistic uniform $\lambda_{\mathrm{Edd}}$ distribution for the synthetic AGN populations. In order to account for a realistic distribution of $\lambda_{\mathrm{Edd}}$ we used the 9-months {\it Swift}/BAT AGN sample. We took the values of the 2--10\,keV luminosity and of $M_{\mathrm{BH}}$ reported in the X-ray study of \citet{Winter:2009uq}. We assumed a constant 2--10\,keV bolometric correction of $\kappa_{\mathrm{x}}=20$, and took into account the 52 objects with $\log \lambda_{\mathrm{Edd}}\geq -2$. For each geometry we used the set of parameters for which the steepest slope $B$ was obtained (see Sect.\,\ref{sect:toroidal}-\ref{sect:slab_geometry}), and ran 1,000 Monte-Carlo simulations using Eq.\,\ref{eq:lambda_gamma} to obtain $\Gamma$, and assuming a random inclination angle between $1^{\circ}$ and $\theta_{\mathrm{OA}}$ for the spherical-toroidal geometry, and between $0^{\circ}$ and $60^{\circ}$ for the remaining two geometries.  We obtained an average value of $B=-0.080$ with a standard deviation of $\sigma_{\rm\,B}=0.005$ for the toroidal geometry, while $B=-0.048$ with $\sigma_{\rm\,B}=0.015$ for the spherical-toroidal geometry, and $B=-0.051$ with $\sigma_{\rm\,B}=0.016$ for the slab geometry.

To account for the possible error introduced by using constant bolometric corrections we considered the results reported by \citet{Vasudevan:2010kx} for the 9-months {\it Swift}/BAT AGN sample. In their work they calculated the values of $\kappa_{\mathrm{x}}$ using the {\it Swift}/BAT and {\it IRAS} fluxes together with the AGN nuclear SED templates of \citet{Silva:2004vn}. We considered only the 28 unobscured AGN for which the values of $\kappa_{\mathrm{x}}$ were available, in order to have a sample consistent with those usually used for studies of the X-ray Baldwin effect. Following what we did for the sample of \citet{Winter:2009uq}, we obtained from our simulations an average value of $B=-0.084$ with a standard deviation of $\sigma_{\rm\,B}=0.009$ for the toroidal geometry, $B=-0.037$ with $\sigma_{\rm\,B}=0.032$ for the spherical-toroidal geometry, and $B=-0.050$ with $\sigma_{\rm\,B}=0.033$ for the slab geometry. The larger dispersion associated with these simulations with respect to those performed using the sample of \citet{Winter:2009uq} is likely related to the lower number of objects we used. Due to the larger scatter obtained, for the toroidal geometry $\sim 3\%$ of the of the Monte-Carlo runs produces values of $B$ consistent within 1$\sigma$ with {\it Chandra}/HEG observations, $\sim 2\%$ for the spherical-toroidal geometry and $\sim 7\%$ for the slab geometry.

As a last test we used the Eddington ratio distribution obtained from the Hamburg/ESO survey of type-I low-redshift ($z<0.3$) AGN by \citet{Schulze:2010uq}. In their work they found that, after correcting for observational biases, the intrinsic $\lambda_{\mathrm{Edd}}$ distribution of unobscured AGN can be well represented by a log-normal distribution with a mean of $\log \lambda_{\mathrm{Edd}}=-1.83$ and a standard deviation of $\sigma_{\lambda}=0.49$. Generating random values of $\lambda_{\mathrm{Edd}}$ distributed with a log-normal distribution (for $\log \lambda_{\mathrm{Edd}}> -2$) with the parameters found by \citet{Schulze:2010uq}, we created a dummy AGN population of 100 sources. We used this population to perform 1,000 Monte-Carlo simulations following what we did for the samples of \citet{Winter:2009uq} and \citet{Vasudevan:2010kx}. We obtained an average value of $B=-0.090$ with a standard deviation of $\sigma_{\rm\,B}=0.008$ for the toroidal geometry, $B=-0.041$ with $\sigma_{\rm\,B}=0.028$ for the spherical-toroidal geometry, and $B=-0.040$ with $\sigma_{\rm\,B}=0.030$ for the slab geometry. For the toroidal geometry $\sim12\%$ of the simulations produce values of $B$ consistent with the observations, while lower percentages are obtained for the spherical-toroidal ($\sim 3\%$) and slab ($\sim 2\%$) geometries.

\begin{figure}
\begin{center}
\includegraphics[width=8.7cm]{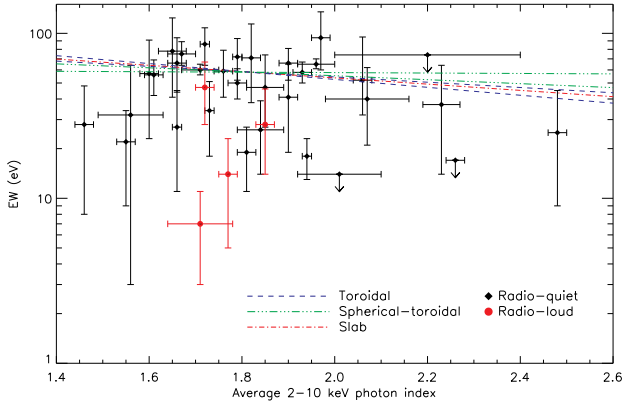}
\caption{Scatter plot of the EW of the narrow component of the iron K$\alpha$ line versus the photon index for the {\it Chandra}/HEG sample. The values of EW are those reported in \citet{Shu:2010zr} fixing $\sigma=1\rm\,eV$ and averaging when possible different observations of the same source. The values of $\Gamma$ were taken from the literature (see Appendix\,\ref{list_sources}), averaging all the results of the last $\sim10$\,years obtained with {\it Chandra}, {\it XMM-Newton} and {\it Suzaku}. The lines represent the $EW-\Gamma$ trends expected considering three different geometries of the reprocessing material: toroidal (blue dashed lines), spherical-toroidal (green dot-dot-dashed line) and slab (red dot-dashed line). The values of $\alpha$ were obtained fitting the {\it Chandra}/HEG data, fixing $\beta$ to the results of the simulations (see Eq.\,\ref{eq:ewvsgamma}). For the toroidal and the spherical-toroidal geometries the values of $\beta$ used are the maximum and minimum obtained by the simulations.}
\label{fig:iron_gamma}
\end{center}
\end{figure}

\section{The relation between the iron K$\alpha$ EW and $\Gamma$}\label{Sect:EWvsGamma}
So far we have discussed the influence of the dependence of $\Gamma$ on $\lambda_{\rm\,Edd}$ on the X-ray Baldwin effect. In this section we investigate the relation between the narrow component of the Fe K$\alpha$ EW and the photon index. A clear trend between these two quantities might in fact suggest that the $\Gamma-\lambda_{\rm\,Edd}$ correlation plays an important role in the observed $EW-\lambda_{\mathrm{Edd}}$ relation. Using {\it ASCA} observations of 25 unabsorbed AGN, \citet{Lubinski:2001kx} found evidence of a possible positive trend between the iron K$\alpha$ EW and $\Gamma$. A weak positive correlation was also found by \citet{Perola:2002vn} studying {\it BeppoSAX} observations of bright AGN. A more complex relationship between these two quantities was found by the {\it RXTE} study of \citet{Mattson:2007bh}. In their study they analysed 350 time-resolved spectra of 12 type-I AGN, and found that the EW of the iron line is positively correlated with the photon index up to $\Gamma \simeq 2$, where the trend turns over and the two parameters become anti-correlated. However, in all these studies the EW might also include the contribution of a broad component. 

\subsection{The {\it Chandra}/HEG sample}
In order to study the relation between the narrow component of the iron K$\alpha$ line and the photon index we used the EW obtained by {\it Chandra}/HEG observations of a sample of 36 type-1/1.5 AGN (fixing $\sigma=1\rm\,eV$) and reported in \citet{Shu:2010zr}. To reduce the effects of variability of the primary continuum, we used the values of EW averaged, when possible, over several observations. We collected the values of $\Gamma$ from {\it XMM-Newton}, {\it Chandra} and {\it Suzaku} studies of the sources performed in the last ten years (see Appendix \ref{list_sources}), and used the average of all the observations, in order to reduce the impact of photon-index variability. The material responsible for the narrow component of the Fe K$\alpha$ line is in fact thought to be located at several light years from the central engine, so that the response of the line to changes of $\Gamma$ is not simultaneous. The scatter plot of the EW of the narrow component of the iron K$\alpha$ line versus the photon index is shown in Fig.\,\ref{fig:iron_gamma}. From the figure one can see that the data are dominated by scatter, and do not appear to be neither linearly correlated nor constant, with the Spearman's rank correlation coefficient being $\rho=-0.01$. Excluding the upper limits and fitting the data with a constant, one obtains $\chi^{2}\simeq 73$ for 32 degrees of freedom. Using the \texttt{fitexy} procedure to fit the data with
\begin{equation}\label{eq:ewvsgamma}
\log EW = \alpha + \beta\,\Gamma,
\end{equation}
we obtained $\alpha=1.8\pm0.2$ and $\beta=-0.05\pm0.12$.

\begin{figure}
\begin{center}
\includegraphics[width=8.7cm]{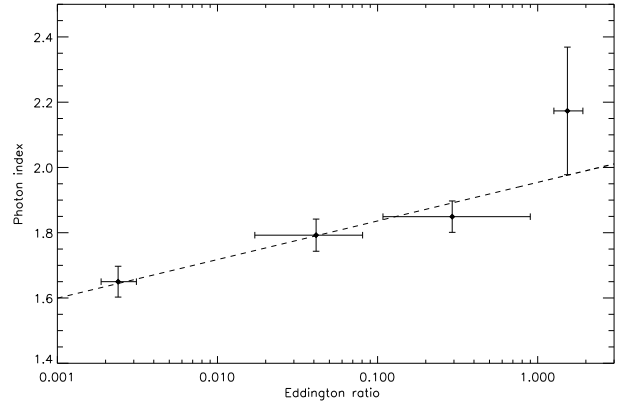}
\caption{Rebinned scatter plot of the photon index versus the Eddington ratio of the {\it Chandra}/HEG type-I AGN sample of \citet{Shu:2010zr}. The dashed line represents the best fit to the data ($\Gamma\propto 0.12\log \lambda_{\mathrm{Edd}}$).}
\label{fig:edd_gammaC}
\end{center}
\end{figure}

To compare the data to the expected $EW-\Gamma$ trend, we investigated the relation obtained by the three different geometries. This was done simulating, similarly to what was done in Sect.\,\ref{sect:simulations}, several unabsorbed populations of AGN, with a uniform distribution of $\Gamma$ between 1.5 and 2.5. We fitted the simulated data with Eq.\,\ref{eq:ewvsgamma}, using the weighted least-square method, with the same weights we used for Eq.\,\ref{Eq:fit}. In the case of the toroidal geometry $\beta$ varies between $-0.16$ (for $\log N_{\rm\,H}^{\mathrm{\,T}}=22.5$) and $-0.23$ (for $\log N_{\rm\,H}^{\mathrm{\,T}}=25$). For the spherical toroidal geometry the values of $\beta$ vary between $ -0.09$ (for $\theta_{\mathrm{OA}}=70^{\circ}$ and $\log N_{\rm\,H}^{\mathrm{\,T}}=22.6$) and $ -0.15$ (for $\theta_{\mathrm{OA}}=10^{\circ}$ and $\log N_{\rm\,H}^{\mathrm{\,T}}=24.5$), while for the slab geometry we obtained $\beta=-0.28$ (for $R=1$). The expected trends are plotted in Fig.\,\ref{fig:iron_gamma}. Given the heterogeneous sampling of the values of $\Gamma$ and the large scatter in the data, we cannot conclude whether the photon index plays a dominant role or not.

In Fig.\,\ref{fig:edd_gammaC} we plot the rebinned scatter plot of the photon index versus the Eddington ratio for the {\it Chandra}/HEG sample. Fitting the data with a log-linear function we obtained a dependence of $\Gamma$ on $\log \lambda_{\mathrm{Edd}}$ significantly weaker than the one found by \citet{Shemmer:2008fk}:
\begin{equation}\label{eq:Gamm_eddratio_Chandra}
\Gamma= (1.95\pm0.06) + (0.12\pm0.04) \log \lambda_{\mathrm{Edd}}.
\end{equation}
However, the intrinsic correlation might at least in part be attenuated by the fact that we considered values of $\Gamma$ which were averaged over the last $\sim 10$\,years, while the values of the Eddington ratio were obtained using only {\it Chandra}/HEG observations. The photon index is in fact expected to react on short time scales to changes of $\lambda_{\mathrm{Edd}}$, so that our treatment is likely to introduce additional scatter to the correlation.

\subsection{Reducing the effects of variability using {\it Suzaku} and {\it Swift}/BAT}
In order to reduce the possible effects of variability we expanded our study using the {\it Suzaku} sample of \citet{Fukazawa:2011fk} and the results of the 58-months 14--195\,keV {\it Swift}/BAT catalog \citep{Baumgartner:2010uq}. The flux of the continuum is more variable than that of the Fe K$\alpha$ line, and using the 58-months averaged 14--195\,keV {\it Swift}/BAT flux ($F_{\mathrm{BAT}}$) allows us to smooth out the effects of continuum variability. To have a proxy of the Fe K$\alpha$ EW less sensitive to variability we divided the flux of the iron K$\alpha$ line ($F_{\mathrm{K}\alpha}$) obtained by {\it Suzaku} by $F_{\mathrm{BAT}}$, and studied the relation of $F_{\mathrm{K}\alpha}/F_{\mathrm{BAT}}$ with the {\it Swift}/BAT photon index. Of the 87\,AGN\footnote{In \citet{Fukazawa:2011fk} a total of 88 AGN are reported, but NGC\,1142 and Swift\,J0255.2$-$0011 are the same source.} reported in the work of \citet{Fukazawa:2011fk} we excluded the 13 Compton-thick Seyfert\,2s because their BAT flux is likely to be significantly attenuated, the seven objects for which no iron K$\alpha$ line was detected by {\it Suzaku} and the four sources not detected by {\it Swift}/BAT. The final sample we used contained 63 AGN, of which 30 Seyfert\,1s, 31 Seyfert\,2s and 2 LINERs. For the 11 objects for which several {\it Suzaku} observations were reported we used the average values of the iron K$\alpha$ flux. In Fig.\,\ref{Fig:ratiovsGamma} the scatter plot of $F_{\mathrm{K}\alpha}$/$F_{\mathrm{BAT}}$ versus the {\it Swift}/BAT photon index is shown. Only a moderately significant positive correlation between the two quantities is found ($\rho=0.3$, $P_{\mathrm{n}}\sim 2\%$). We fitted the data with a linear relationship of the type $\log(F_{\mathrm{K}\alpha}$/$F_{\mathrm{BAT}})=\xi+\mu\,\Gamma$. Using the \texttt{fitexy} procedure we obtained a slope of $\mu=0.58\pm0.13$. A positive correlation is at odds with what would be expected if EW and $\Gamma$ were strongly related. However, the sample we used includes both type-I and type-II AGN; considering only type-I objects, consistently to what is usually done for the X-ray Baldwin effect, the correlation is not significant anymore ($P_{\mathrm{n}}=54\%$).

\begin{figure}
\begin{center}
\includegraphics[width=\columnwidth]{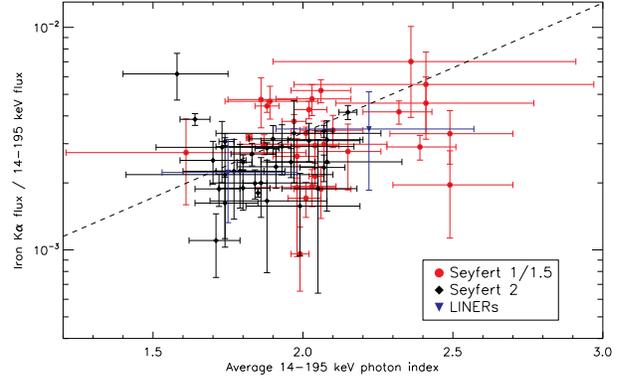}

\caption{Ratio between the iron K$\alpha$ line flux measured by {\it Suzaku} \citep{Fukazawa:2011fk} and the 14--195\,keV {\it Swift}/BAT flux \citep{Baumgartner:2010uq} versus the {\it Swift}/BAT photon index. The dashed line represents the linear fit to the data obtained using the \texttt{fitexy} procedure.}
\label{Fig:ratiovsGamma}
\end{center}
\end{figure}

\section{Discussion}\label{sect:discussion}

In the last years several works have found a clear correlation between the X-ray photon index and the Eddington ratio (e.g., \citealp{Shemmer:2008fk}, \citealp{Fanali:2013fk}). Discordant results have been obtained when looking for a correlation between the X-ray luminosity and the photon index. While some works found an anti-correlation (e.g., \citealp{Green:2009fk,Young:2009uq,Corral:2011kx}), others found a positive correlation (e.g., \citealp{Dai:2004zr,Saez:2008ys}), and many did not find any trend between the two parameters (e.g., \citealp{Nandra:1994ly,George:2000hc,Reeves:2000uq,Page:2005qf,Vignali:2005bh,Shemmer:2005dq,Shemmer:2006uq,Just:2007cr,Risaliti:2009nx}). For individual AGN several works found a tight correlation between $\Gamma$ and $L_{\,2-10}$ (e.g., \citealp{Magdziarz:1998kl,Zdziarski:2003tg,Sobolewska:2009kx}), which also points towards $\lambda_{\mathrm{Edd}}$ being the main driver of the observed trend. Changes in luminosity for individual objects are in fact directly related to changes of the Eddington ratio. In this work we have studied the influence of the $\Gamma-\lambda_{\mathrm{Edd}}$ relation on the $EW-\lambda_{\mathrm{Edd}}$ trend for different geometries of the reflecting material. We have found that assuming $\Gamma\propto 0.31\log \lambda_{\mathrm{Edd}}$ as found by \citet{Shemmer:2008fk} it is not possible to completely reproduce the observed $EW-\lambda_{\mathrm{Edd}}$ correlation, with the average slope produced being at most $B\simeq -0.08$ for the toroidal geometry, $B\simeq -0.04$ for the spherical toroidal geometry and $B\simeq -0.05$ for the slab geometry. 

Recently the $\Gamma-\lambda_{\mathrm{Edd}}$ correlation has been confirmed by the study of \citet{Risaliti:2009nx} on a large sample of $\sim 400$ Sloan Digital Sky Survey quasars with available {\it XMM-Newton} X-ray spectra. In their work \citet{Risaliti:2009nx} found a trend consistent with that observed by \citet{Shemmer:2008fk} when studying objects with black hole masses estimates obtained using $\mathrm{H}\beta$, Mg\,II and C\,IV. When considering only the $\sim80$ objects with values of $M_{\mathrm{BH}}$ obtained with $\mathrm{H}\beta$ the correlation becomes stronger ($\Gamma \propto 0.58 \log \lambda_{\mathrm{Edd}}$), while it is weaker or absent using only objects with Mg\,II and C\,IV measurements, respectively. The existence of differences between these trends has been interpreted by \citet{Risaliti:2009nx} as being due to the different uncertainties on $M_{\mathrm{BH}}$ obtained from the three lines. While C\,IV is believed to be a poor estimator of the mass of the supermassive black hole (e.g., \citealp{Netzer:2007vn}), both H$\beta$ and Mg\,II are thought to be produced in virialised gas and thus expected to be good mass indicators. However Mg\,II is a doublet and can be affected by contamination of Fe\,II, so that H$\beta$ is likely to be the best mass indicator. If, as argued by \citet{Risaliti:2009nx}, the $\Gamma-\lambda_{\mathrm{Edd}}$ trend obtained using black-hole mass estimates from H$\beta$ is the most reliable, then the influence on the X-ray Baldwin effect should be more important. A result consistent with that of \citet{Risaliti:2009nx} has been obtained by \citet{Jin:2012fk} using black-hole mass estimates obtained from H$\alpha$ and H$\beta$, while \citet{Brightman:2013vn} using H$\alpha$ and Mg\,II found a flatter slope ($0.32\pm0.05$). Performing the same simulations of Sect.\,\ref{sect:toroidal}-\ref{sect:slab_geometry} using the $\Gamma-\lambda_{\mathrm{Edd}}$ correlation found by \citet{Risaliti:2009nx} using H$\beta$, we obtained values of $B$ between $-0.093$ (for $\log N_{\rm\,H}^{\mathrm{\,T}}=22.5$) and $-0.14$ (for $\log N_{\rm\,H}^{\mathrm{\,T}}=25$) for the toroidal geometry, between $-0.008$ (for $\theta_{\mathrm{OA}}=70^{\circ}$ and $\log N_{\rm\,H}^{\mathrm{\,T}}=22.6$) and $ -0.07$ (for $\theta_{\mathrm{OA}}=10^{\circ}$ and $\log N_{\rm\,H}^{\mathrm{\,T}}=24.5$) for the spherical-toroidal geometry, and of $-0.11$ (for $R=1$) for the slab geometry. Repeating the Monte-Carlo simulations of Sect.\,\ref{Sect:edd_distr} adopting the $\lambda_{\mathrm{Edd}}$ distribution of \citet{Vasudevan:2010kx} and using the results of \citet{Risaliti:2009nx} we found an average value of $B\simeq-0.147$ ($\sigma=0.009$) for the toroidal geometry, $B\simeq -0.09$ ($\sigma=0.04$) for the spherical-toroidal geometry, and of $B\simeq -0.09$ ($\sigma=0.03$) for the slab geometry. Thus for all the geometries, assuming this steeper $\Gamma-\lambda_{\mathrm{Edd}}$ relation one obtains slopes consistent within 1$\sigma$ with the time-averaged (i.e. fits per source) $EW-\lambda_{\mathrm{Edd}}$ trend.

\citet{Gu:2009uq}, studying a sample of low luminosity AGN (LLAGN), which included low-ionisation nuclear emission line regions (LINERs) and local Seyfert galaxies with $\log \lambda_{\mathrm{Edd}} \lesssim -2$, found that, contrarily to what has been observed at higher values of Eddington ratios, for these objects the photon index appears to be anti-correlated to $\lambda_{\mathrm{Edd}}$. They found that $\Gamma \propto (-0.09\pm 0.03)\log \lambda_{\mathrm{Edd}}$, and argued that, as the anti-correlation is consistent with that found for X-ray binaries (XRB) in a low/hard state (e.g., \citealp{Yamaoka:2005ys}), LLAGN and XRB in low/hard state might have a similar accretion mechanism, possibly an advection-dominated accretion flow (ADAF; see also \citealp{Qiao:2012uq}). A stronger anti-correlation has been found by \cite{Younes:2011cr} studying a sample of 13 LINERs ($\Gamma \propto -0.31\,\log \lambda_{\mathrm{Edd}}$). These results might be used to constrain the importance of the $\Gamma-\lambda_{\mathrm{Edd}}$ correlation on the X-ray Baldwin effect. The detection of a significant flattening or of a positive correlation between EW and $\lambda_{\mathrm{Edd}}$ for $\log \lambda_{\mathrm{Edd}} \lesssim -2$ would in fact point towards an important impact on the observed $EW-\lambda_{\mathrm{Edd}}$ relation.

In order to understand which is the mechanism responsible for the X-ray Baldwin effect it is fundamental to assess whether its main driver is the luminosity or rather the Eddington ratio. Our analysis of the {\it Chandra}/HEG sample of \citet{Shu:2010zr} seems to indicate that the correlation of EW with the X-ray luminosity is stronger than that with the Eddington ratio. This would point towards a marginal role of the $\Gamma-\lambda_{\mathrm{Edd}}$ trend on the decline of the iron K$\alpha$ EW with the luminosity and the Eddington ratio, although the significant uncertainties associated to $\kappa_{\mathrm{x}}$ and $M_{\rm\,BH}$ do not allow us to reach a firm conclusion.  A possible alternative explanation for the X-ray Baldwin effect is that it is related to the decrease of the covering factor of the torus with the luminosity. Such a trend has been observed in a large number of studies performed at different wavelengths (e.g., \citealp{Ueda:2003qf}, \citealp{Beckmann:2009fk}), and it is likely to be the cause of the different luminosity distributions of type-I and type-II AGN (e.g., \citealp{Ricci:2011zr}). In a recent work \citep{Ricci:2013uq}, simulating the X-ray spectra of type-I populations with luminosity-evolving tori, we have shown that such an effect is able to reproduce the slope of the X-ray Baldwin effect for a large range of values of equatorial column densities of the torus ($\log N_{\mathrm{H}}^{\mathrm{T}}\geq 23.1$). In order to discern between these two explanations, it is necessary to understand whether the slope of the $\Gamma-\lambda_{\mathrm{Edd}}$ relation is as steep as $\simeq -0.6$ (e.g., \citealp{Risaliti:2009nx}) or rather flatter ($\simeq -0.3$), as shown by several works  (e.g., \citealp{Shemmer:2008fk}).

\section{Summary and conclusions}\label{sect:conclusions}
In this work we have studied the influence of the $\Gamma-\lambda_{\mathrm{Edd}}$ relation on the observed $EW-\lambda_{\mathrm{Edd}}$ trend, assuming three different geometries for the reprocessing material. Our main results are the following.
\begin{itemize}
\item Applying bolometric corrections we found that the $EW-\lambda_{\mathrm{Edd}}$ trend has a slope of $B=-0.13\pm0.03$ (in the {\it per source} case), and that it appears to be less significant then the $EW-L_{\,2-10}$ relation.
\smallskip
\item Using the $\Gamma-\lambda_{\mathrm{Edd}}$ relation of \citet{Shemmer:2008fk} it is not possible to fully account for the X-ray Baldwin effect for none of the geometries considered here, the slopes produced being flatter than the observed one.
\smallskip
\item The relation between EW and $\Gamma$ is dominated by scatter, and no clear anti-correlation is observed.
\smallskip
\item Using the {\it Chandra}/HEG sample we found a weaker dependence of the photon index on the Eddington ratio ($\Gamma\propto 0.12\log \lambda_{\mathrm{Edd}}$) than that obtained by \citet{Shemmer:2008fk}. This might be related to the additional scatter introduced by the different sampling we used for $\lambda_{\rm\,Edd}$ and $\Gamma$.
\smallskip
\item If the recent results of \citet{Risaliti:2009nx} and \citet{Jin:2012fk}, who found a steeper correlation between photon index and Eddington ratio than that of \citet{Shemmer:2008fk}, were to be confirmed, then the $\Gamma-\lambda_{\mathrm{Edd}}$ correlation might be able to produce slopes consistent with the X-ray Baldwin effect.
\end{itemize}

Our findings show that a good knowledge of the slope of the $\Gamma-\lambda_{\mathrm{Edd}}$ relation is critical to understand whether this trend plays a leading or rather a marginal role in the X-ray Baldwin effect. A relation as steep as that found by \citet{Risaliti:2009nx} would point towards a leading role, while a flatter trend, as that obtained by \citet{Shemmer:2008fk}, would imply that the X-ray Baldwin effect is driven by another mechanism. A plausible candidate would be the decrease of the covering factor of the molecular torus with the luminosity, which, as recently shown in \citet{Ricci:2013uq}, can straightforwardly explain the observed decline of the iron K$\alpha$ EW with the luminosity.

\appendix

\section{$\Gamma-EW$ data}\label{list_sources}
In Table\,\ref{tab:data} the values of the average photon index of the {\it Chandra}/HEG sample of \citet{Shu:2010zr} are listed, together with their references.

\begin{table*}
\caption[]{(1) Average values and (2) references of the photon index of the objects in the {\it Chandra}/HEG sample of \citet{Shu:2010zr}.}
\label{tab:data}
\begin{center}
\begin{tabular}{lcc}
\hline
\hline
\noalign{\smallskip}
& (1) & (2)      \\ 
\noalign{\smallskip}
Source 						 & $<\Gamma>$  & References      \\ 
\noalign{\smallskip}
\hline
\noalign{\smallskip}
Fairall 9 	    					 	& $1.85\pm0.04$  & \citet{Gondoin:2001fk,Emmanoulopoulos:2011fk,Patrick:2011zt,McKernan:2007bh} \\ 
\noalign{\smallskip}
NGC 526a	   			 	&  $1.46\pm0.02$   & \citet{Guainazzi:2011kx}  \\
\noalign{\smallskip}
Mrk 590     					 	&    $1.65\pm0.03$ & \citet{Gallo:2006vn,Longinotti:2007ys} \\ 
\noalign{\smallskip}
NGC 985     					 	&    $1.60\pm0.02$ & \citet{Krongold:2005jl,Krongold:2009zr} \\
\noalign{\smallskip}
ESO 198$-$G24  			 	&   $1.82\pm0.02$   & \citet{Porquet:2004ly,Winter:2012ve}  \\ 
\noalign{\smallskip}
3C 120    						 	&   $1.72\pm0.02$ & \citet{McKernan:2007bh,Miyazawa:2009kl,Brightman:2011qf} \\ 
\noalign{\smallskip}
NGC 2110    			         &  $1.67\pm0.03$  & \citet{Evans:2007bh,Miyazawa:2009kl}  \\  
\noalign{\smallskip}
PG 0844$+$349  			 &  $2.06\pm0.02$   & \citet{Gallo:2011dq} \\ 
\noalign{\smallskip}
Mrk 705     					 	&  $2.2\pm 0.2$    &  \citet{Gallo:2005cr} \\ 
\noalign{\smallskip}
MCG $-$5$-$23$-$16 			 	&  $1.79\pm0.02$ & \citet{Dewangan:2003kl,Braito:2007nx,Reeves:2007oq,Guainazzi:2011kx} \\ 
\noalign{\smallskip}
NGC 3227    					 	&  $1.56 \pm 0.07$  & \citet{Gondoin:2003hc,Markowitz:2009tg}   \\
\noalign{\smallskip}
NGC 3516    				 	&  $1.93 \pm0.02$  & \citet{Turner:2005ij,Markowitz:2008fv,Turner:2011bs} \\
\noalign{\smallskip}
NGC 3783    				 	&  $1.71\pm0.04$ & \citet{Kaspi:2001fu,Blustin:2002qa,Netzer:2003kl};  \\
 	& &  \citet{Reeves:2004mi,Reis:2012pi} \\
\noalign{\smallskip}
NGC 4051   				 	&   $1.97\pm0.02$ & \citet{Ponti:2006uq,Lobban:2011fk} \\
\noalign{\smallskip}
NGC 4151   				 	&   $1.96\pm0.04$ & \citet{Cappi:2006ys,Lubinski:2010kx,Wang:2010vn} \\
\noalign{\smallskip}
Mrk 766     					 	&  $2.23\pm0.04$ & \citet{Boller:2001zr,Turner:2007ve,Risaliti:2011ly} \\
\noalign{\smallskip}
3C 273      				 	&   $1.71\pm 0.07$   & \citet{Soldi:2008qf} \\
\noalign{\smallskip}
NGC 4593    				 	&  $1.76\pm0.02$   & \citet{Steenbrugge:2003cr,Brenneman:2007dq,McKernan:2007bh};  \\
 			     & &\citet{Markowitz:2009nx}  \\
\noalign{\smallskip}
MCG $-$6$-$30$-$15 		 & $1.94\pm0.01$   & \citet{Wilms:2001tg,Lee:2002kl,Fabian:2002hc,Chiang:2011oq} \\
\noalign{\smallskip}
IRAS 13349$+$2438		 &	 $2.07\pm0.09$  & \citet{Longinotti:2003bs,Holczer:2007ij}  \\
\noalign{\smallskip}
IC 4329A   					 	&  $1.81\pm0.02$   & \citet{Gondoin:2001dz,McKernan:2004fv,Steenbrugge:2005fu,Miyazawa:2009kl}  \\
\noalign{\smallskip}
Mrk 279     					 	&   $1.66\pm0.02$ & \citet{McKernan:2007bh,Costantini:2010qa} \\
\noalign{\smallskip}
NGC 5506    					 	&  $1.90\pm0.02$ & \citet{Bianchi:2003mi,Miyazawa:2009kl,Guainazzi:2010pi}  \\
\noalign{\smallskip}
NGC 5548     				 	&   $1.61\pm0.02$ & \citet{Andrade-Velazquez:2010uq,de-La-Calle-Perez:2010vn,Liu:2010ys} \\
\noalign{\smallskip}
Mrk 290     				 	&  $1.66\pm0.01$  & \citet{Zhang:2011zr} \\
\noalign{\smallskip}
PDS 456     					 	&  $2.26\pm0.02$  & \citet{Reeves:2003ve,Reeves:2009qf,Behar:2010ly}  \\
\noalign{\smallskip}
E1821$+$643      				 	&   $1.84\pm0.05$ & \citet{Fang:2002dq,Jimenez-Bailon:2007cr,Russell:2010bh} \\
\noalign{\smallskip}
3C 382     			 	&    $1.77\pm0.02$  & \citet{Gliozzi:2007nx,Sambruna:2011oq}  \\
\noalign{\smallskip}
IRAS 18325$-$5926		 		&  $2.01\pm0.09$   & \citet{Mocz:2011kl} \\
\noalign{\smallskip}
4C $+$74.26    			 	&   $1.85\pm0.02$  & \citet{Ballantyne:2005tg,Larsson:2008hc}  \\
\noalign{\smallskip}
Mrk 509     					 	&   $1.73\pm0.01$ & \citet{Yaqoob:2003ij,Ponti:2009bs,Noda:2011fv,Petrucci:2013fk} \\
\noalign{\smallskip}
NGC 7213    			        & $1.72\pm0.01$ & \citet{Starling:2005fu,Bianchi:2008dz,Lobban:2010kl} \\
\noalign{\smallskip}
NGC 7314    				 	&  $1.90\pm0.02$ & \citet{Yaqoob:2003qa,Ebrero:2011mi} \\
\noalign{\smallskip}
Ark 564     					 	& $2.48\pm0.02$ & \citet{Matsumoto:2004pi,Vignali:2004ff,Miyazawa:2009kl,Singh:2011lh} \\
\noalign{\smallskip}
MR 2251$-$178     				 	& $1.55\pm 0.02$  & \citet{Gibson:2005fu,Ramirez:2008fk} \\
\noalign{\smallskip}
NGC 7469     				 	&   $1.79\pm0.01$ & \citet{Blustin:2003tw,Scott:2005qo,Blustin:2007il,Patrick:2011zt} \\
\noalign{\smallskip}
\hline
\noalign{\smallskip}
\end{tabular}        
\end{center}
\end{table*}

\section*{Acknowledgements}
We thank R. Sato for his help, Chin Shin Chang and Poshak Gandhi for their comments on the manuscript. CR thanks P.O. Petrucci, the Sherpa group and IPAG for hospitality during his stay in Grenoble. We thank the anonymous referee for his/her comments that helped improving the paper. CR is a Fellow of the Japan Society for the Promotion of Science (JSPS). This work was partly supported by the Grant-in-Aid for Scientific Research 23540265 (YU) from the Ministry of Education, Culture, Sports, Science and Technology of Japan (MEXT). This research has made use of the NASA/IPAC Extragalactic Database (NED) which is operated by the Jet Propulsion Laboratory, of the High Energy Astrophysics Science Archive Research Center (HEASARC), provided by NASA's Goddard Space Flight Center, and of the SIMBAD Astronomical Database which is operated by the Centre de Donn\'ees astronomiques de Strasbourg.  
 
\bibliographystyle{mnras}
 \bibliography{lambda_gamma}

\begin{thebibliography}{166}
\expandafter\ifx\csname natexlab\endcsname\relax\def\natexlab#1{#1}\fi

\bibitem[{Anders} \& {Ebihara}(1982)]{Anders:1982uq}
{Anders} E., {Ebihara} M., 1982, \gca, 46, 2363

\bibitem[{Andrade-Vel{\'a}zquez} et~al.(2010){Andrade-Vel{\'a}zquez},
  {Krongold}, {Elvis} et~al.]{Andrade-Velazquez:2010uq}
{Andrade-Vel{\'a}zquez} M., {Krongold} Y., {Elvis} M., et~al., 2010, \apj, 711,
  888

\bibitem[{Arnaud}(1996)]{Arnaud:1996kx}
{Arnaud} K.~A., 1996, in { Astronomical Data Analysis Software and Systems
  V\/}, edited by {G.~H.~Jacoby \& J.~Barnes}, vol. 101 of { Astronomical
  Society of the Pacific Conference Series\/}, ~17

\bibitem[{Baldwin}(1977)]{Baldwin:1977fk}
{Baldwin} J.~A., 1977, \apj, 214, 679

\bibitem[{Ballantyne} \& {Fabian}(2005)]{Ballantyne:2005tg}
{Ballantyne} D.~R., {Fabian} A.~C., 2005, \apjl, 622, L97

\bibitem[{Baumgartner} et~al.(2010){Baumgartner}, {Tueller}, {Markwardt} \&
  {Skinner}]{Baumgartner:2010uq}
{Baumgartner} W.~H., {Tueller} J., {Markwardt} C., {Skinner} G., 2010, in {
  AAS/High Energy Astrophysics Division \#11\/}, vol.~42 of { Bulletin of the
  American Astronomical Society\/},  675

\bibitem[{Beckmann} et~al.(2009){Beckmann}, {Soldi}, {Ricci}
  et~al.]{Beckmann:2009fk}
{Beckmann} V., {Soldi} S., {Ricci} C., et~al., 2009, \aap, 505, 417

\bibitem[{Behar} et~al.(2010){Behar}, {Kaspi}, {Reeves}, {Turner}, {Mushotzky}
  \& {O'Brien}]{Behar:2010ly}
{Behar} E., {Kaspi} S., {Reeves} J., {Turner} T.~J., {Mushotzky} R., {O'Brien}
  P.~T., 2010, \apj, 712, 26

\bibitem[{Bian}(2005)]{Bian:2005ve}
{Bian} W.-H., 2005, Chinese Journal of Astronomy and Astrophysics Supplement,
  5, 289

\bibitem[{Bianchi} et~al.(2003){Bianchi}, {Balestra}, {Matt}, {Guainazzi} \&
  {Perola}]{Bianchi:2003mi}
{Bianchi} S., {Balestra} I., {Matt} G., {Guainazzi} M., {Perola} G.~C., 2003,
  \aap, 402, 141

\bibitem[{Bianchi} et~al.(2007){Bianchi}, {Guainazzi}, {Matt} \& {Fonseca
  Bonilla}]{Bianchi:2007vn}
{Bianchi} S., {Guainazzi} M., {Matt} G., {Fonseca Bonilla} N., 2007, \aap, 467,
  L19

\bibitem[{Bianchi} et~al.(2008){Bianchi}, {La Franca}, {Matt}
  et~al.]{Bianchi:2008dz}
{Bianchi} S., {La Franca} F., {Matt} G., et~al., 2008, \mnras, 389, L52

\bibitem[{Blustin} et~al.(2002){Blustin}, {Branduardi-Raymont}, {Behar}
  et~al.]{Blustin:2002qa}
{Blustin} A.~J., {Branduardi-Raymont} G., {Behar} E., et~al., 2002, \aap, 392,
  453

\bibitem[{Blustin} et~al.(2003){Blustin}, {Branduardi-Raymont}, {Behar}
  et~al.]{Blustin:2003tw}
{Blustin} A.~J., {Branduardi-Raymont} G., {Behar} E., et~al., 2003, \aap, 403,
  481

\bibitem[{Blustin} et~al.(2007){Blustin}, {Kriss}, {Holczer}
  et~al.]{Blustin:2007il}
{Blustin} A.~J., {Kriss} G.~A., {Holczer} T., et~al., 2007, \aap, 466, 107

\bibitem[{Boller} et~al.(1996){Boller}, {Brandt} \& {Fink}]{Boller:1996kl}
{Boller} T., {Brandt} W.~N., {Fink} H., 1996, \aap, 305, 53

\bibitem[{Boller} et~al.(2001){Boller}, {Keil}, {Tr{\"u}mper}, {O'Brien},
  {Reeves} \& {Page}]{Boller:2001zr}
{Boller} T., {Keil} R., {Tr{\"u}mper} J., {O'Brien} P.~T., {Reeves} J., {Page}
  M., 2001, \aap, 365, L146

\bibitem[{Braito} et~al.(2007){Braito}, {Reeves}, {Dewangan}
  et~al.]{Braito:2007nx}
{Braito} V., {Reeves} J.~N., {Dewangan} G.~C., et~al., 2007, \apj, 670, 978

\bibitem[{Brandt} \& {Boller}(1998)]{Brandt:1998kx}
{Brandt} N., {Boller} T., 1998, Astronomische Nachrichten, 319, 7

\bibitem[{Brandt} et~al.(1997){Brandt}, {Mathur} \& {Elvis}]{Brandt:1997fk}
{Brandt} W.~N., {Mathur} S., {Elvis} M., 1997, \mnras, 285, L25

\bibitem[{Brenneman} et~al.(2007){Brenneman}, {Reynolds}, {Wilms} \&
  {Kaiser}]{Brenneman:2007dq}
{Brenneman} L.~W., {Reynolds} C.~S., {Wilms} J., {Kaiser} M.~E., 2007, \apj,
  666, 817

\bibitem[{Brightman} \& {Nandra}(2011)]{Brightman:2011qf}
{Brightman} M., {Nandra} K., 2011, \mnras, 413, 1206

\bibitem[{Brightman} et~al.(2013){Brightman}, {Silverman}, {Mainieri}
  et~al.]{Brightman:2013vn}
{Brightman} M., {Silverman} J.~D., {Mainieri} V., et~al., 2013, \mnras

\bibitem[{Cappi} et~al.(2006){Cappi}, {Panessa}, {Bassani}
  et~al.]{Cappi:2006ys}
{Cappi} M., {Panessa} F., {Bassani} L., et~al., 2006, \aap, 446, 459

\bibitem[{Chiang} \& {Fabian}(2011)]{Chiang:2011oq}
{Chiang} C.-Y., {Fabian} A.~C., 2011, \mnras, 414, 2345

\bibitem[{Corral} et~al.(2011){Corral}, {Della Ceca}, {Caccianiga}
  et~al.]{Corral:2011kx}
{Corral} A., {Della Ceca} R., {Caccianiga} A., et~al., 2011, \aap, 530, A42

\bibitem[{Costantini} et~al.(2010){Costantini}, {Kaastra}, {Korista}
  et~al.]{Costantini:2010qa}
{Costantini} E., {Kaastra} J.~S., {Korista} K., et~al., 2010, \aap, 512, A25

\bibitem[{Dai} et~al.(2004){Dai}, {Chartas}, {Eracleous} \&
  {Garmire}]{Dai:2004zr}
{Dai} X., {Chartas} G., {Eracleous} M., {Garmire} G.~P., 2004, \apj, 605, 45

\bibitem[{de La Calle P{\'e}rez} et~al.(2010){de La Calle P{\'e}rez},
  {Longinotti}, {Guainazzi} et~al.]{de-La-Calle-Perez:2010vn}
{de La Calle P{\'e}rez} I., {Longinotti} A.~L., {Guainazzi} M., et~al., 2010,
  \aap, 524, A50

\bibitem[{De Marco} et~al.(2013){De Marco}, {Ponti}, {Cappi}
  et~al.]{De-Marco:2013fk}
{De Marco} B., {Ponti} G., {Cappi} M., et~al., 2013, \mnras, 431, 2441

\bibitem[{Denney} et~al.(2010){Denney}, {Peterson}, {Pogge}
  et~al.]{Denney:2010kx}
{Denney} K.~D., {Peterson} B.~M., {Pogge} R.~W., et~al., 2010, \apj, 721, 715

\bibitem[{Dewangan} et~al.(2003){Dewangan}, {Griffiths} \&
  {Schurch}]{Dewangan:2003kl}
{Dewangan} G.~C., {Griffiths} R.~E., {Schurch} N.~J., 2003, \apj, 592, 52

\bibitem[{Ebrero} et~al.(2011){Ebrero}, {Costantini}, {Kaastra}, {de Marco} \&
  {Dadina}]{Ebrero:2011mi}
{Ebrero} J., {Costantini} E., {Kaastra} J.~S., {de Marco} B., {Dadina} M.,
  2011, \aap, 535, A62

\bibitem[{Emmanoulopoulos} et~al.(2011){Emmanoulopoulos}, {Papadakis},
  {McHardy}, {Nicastro}, {Bianchi} \& {Ar{\'e}valo}]{Emmanoulopoulos:2011fk}
{Emmanoulopoulos} D., {Papadakis} I.~E., {McHardy} I.~M., {Nicastro} F.,
  {Bianchi} S., {Ar{\'e}valo} P., 2011, \mnras, 415, 1895

\bibitem[{Evans} et~al.(2007){Evans}, {Lee}, {Turner}, {Weaver} \&
  {Marshall}]{Evans:2007bh}
{Evans} D.~A., {Lee} J.~C., {Turner} T.~J., {Weaver} K.~A., {Marshall} H.~L.,
  2007, \apj, 671, 1345

\bibitem[{Fabian} et~al.(2002){Fabian}, {Vaughan}, {Nandra}
  et~al.]{Fabian:2002hc}
{Fabian} A.~C., {Vaughan} S., {Nandra} K., et~al., 2002, \mnras, 335, L1

\bibitem[{Fanali} et~al.(2013){Fanali}, {Caccianiga}, {Severgnini}
  et~al.]{Fanali:2013fk}
{Fanali} R., {Caccianiga} A., {Severgnini} P., et~al., 2013, \mnras, 433, 648

\bibitem[{Fang} et~al.(2002){Fang}, {Davis}, {Lee}, {Marshall}, {Bryan} \&
  {Canizares}]{Fang:2002dq}
{Fang} T., {Davis} D.~S., {Lee} J.~C., {Marshall} H.~L., {Bryan} G.~L.,
  {Canizares} C.~R., 2002, \apj, 565, 86

\bibitem[{Fukazawa} et~al.(2011){Fukazawa}, {Hiragi}, {Mizuno}
  et~al.]{Fukazawa:2011fk}
{Fukazawa} Y., {Hiragi} K., {Mizuno} M., et~al., 2011, \apj, 727, 19

\bibitem[{Gallo} et~al.(2005){Gallo}, {Balestra}, {Costantini}
  et~al.]{Gallo:2005cr}
{Gallo} L.~C., {Balestra} I., {Costantini} E., et~al., 2005, \aap, 442, 909

\bibitem[{Gallo} et~al.(2011){Gallo}, {Grupe}, {Schartel} et~al.]{Gallo:2011dq}
{Gallo} L.~C., {Grupe} D., {Schartel} N., et~al., 2011, \mnras, 412, 161

\bibitem[{Gallo} et~al.(2006){Gallo}, {Lehmann}, {Pietsch}
  et~al.]{Gallo:2006vn}
{Gallo} L.~C., {Lehmann} I., {Pietsch} W., et~al., 2006, \mnras, 365, 688

\bibitem[{George} \& {Fabian}(1991)]{George:1991fk}
{George} I.~M., {Fabian} A.~C., 1991, \mnras, 249, 352

\bibitem[{George} et~al.(2000){George}, {Turner}, {Yaqoob}
  et~al.]{George:2000hc}
{George} I.~M., {Turner} T.~J., {Yaqoob} T., et~al., 2000, \apj, 531, 52

\bibitem[{Gibson} et~al.(2005){Gibson}, {Marshall}, {Canizares} \&
  {Lee}]{Gibson:2005fu}
{Gibson} R.~R., {Marshall} H.~L., {Canizares} C.~R., {Lee} J.~C., 2005, \apj,
  627, 83

\bibitem[{Gliozzi} et~al.(2007){Gliozzi}, {Sambruna}, {Eracleous} \&
  {Yaqoob}]{Gliozzi:2007nx}
{Gliozzi} M., {Sambruna} R.~M., {Eracleous} M., {Yaqoob} T., 2007, \apj, 664,
  88

\bibitem[{Gondoin} et~al.(2001{\natexlab{a}}){Gondoin}, {Barr}, {Lumb},
  {Oosterbroek}, {Orr} \& {Parmar}]{Gondoin:2001dz}
{Gondoin} P., {Barr} P., {Lumb} D., {Oosterbroek} T., {Orr} A., {Parmar} A.~N.,
  2001{\natexlab{a}}, \aap, 378, 806

\bibitem[{Gondoin} et~al.(2001{\natexlab{b}}){Gondoin}, {Lumb}, {Siddiqui},
  {Guainazzi} \& {Schartel}]{Gondoin:2001fk}
{Gondoin} P., {Lumb} D., {Siddiqui} H., {Guainazzi} M., {Schartel} N.,
  2001{\natexlab{b}}, \aap, 373, 805

\bibitem[{Gondoin} et~al.(2003){Gondoin}, {Orr}, {Lumb} \&
  {Siddiqui}]{Gondoin:2003hc}
{Gondoin} P., {Orr} A., {Lumb} D., {Siddiqui} H., 2003, \aap, 397, 883

\bibitem[{Green} et~al.(2009){Green}, {Aldcroft}, {Richards}
  et~al.]{Green:2009fk}
{Green} P.~J., {Aldcroft} T.~L., {Richards} G.~T., et~al., 2009, \apj, 690, 644

\bibitem[{Grupe}(2004)]{Grupe:2004fk}
{Grupe} D., 2004, \aj, 127, 1799

\bibitem[{Grupe} et~al.(2010){Grupe}, {Komossa}, {Leighly} \&
  {Page}]{Grupe:2010oq}
{Grupe} D., {Komossa} S., {Leighly} K.~M., {Page} K.~L., 2010, \apjs, 187, 64

\bibitem[{Gu} \& {Cao}(2009)]{Gu:2009uq}
{Gu} M., {Cao} X., 2009, \mnras, 399, 349

\bibitem[{Guainazzi} et~al.(2011){Guainazzi}, {Bianchi}, {de La Calle
  P{\'e}rez}, {Dov{\v c}iak} \& {Longinotti}]{Guainazzi:2011kx}
{Guainazzi} M., {Bianchi} S., {de La Calle P{\'e}rez} I., {Dov{\v c}iak} M.,
  {Longinotti} A.~L., 2011, \aap, 531, A131

\bibitem[{Guainazzi} et~al.(2010){Guainazzi}, {Bianchi}, {Matt}
  et~al.]{Guainazzi:2010pi}
{Guainazzi} M., {Bianchi} S., {Matt} G., et~al., 2010, \mnras, 406, 2013

\bibitem[{Holczer} et~al.(2007){Holczer}, {Behar} \& {Kaspi}]{Holczer:2007ij}
{Holczer} T., {Behar} E., {Kaspi} S., 2007, \apj, 663, 799

\bibitem[{Ikeda} et~al.(2009){Ikeda}, {Awaki} \& {Terashima}]{Ikeda:2009nx}
{Ikeda} S., {Awaki} H., {Terashima} Y., 2009, \apj, 692, 608

\bibitem[{Iwasawa} \& {Taniguchi}(1993)]{Iwasawa:1993ys}
{Iwasawa} K., {Taniguchi} Y., 1993, \apjl, 413, L15

\bibitem[{Jiang} et~al.(2006){Jiang}, {Wang} \& {Wang}]{Jiang:2006vn}
{Jiang} P., {Wang} J.~X., {Wang} T.~G., 2006, \apj, 644, 725

\bibitem[{Jim{\'e}nez-Bail{\'o}n} et~al.(2007){Jim{\'e}nez-Bail{\'o}n},
  {Santos-Lle{\'o}}, {Piconcelli}, {Matt}, {Guainazzi} \&
  {Rodr{\'{\i}}guez-Pascual}]{Jimenez-Bailon:2007cr}
{Jim{\'e}nez-Bail{\'o}n} E., {Santos-Lle{\'o}} M., {Piconcelli} E., {Matt} G.,
  {Guainazzi} M., {Rodr{\'{\i}}guez-Pascual} P., 2007, \aap, 461, 917

\bibitem[{Jin} et~al.(2012){Jin}, {Ward} \& {Done}]{Jin:2012fk}
{Jin} C., {Ward} M., {Done} C., 2012, \mnras, 425, 907

\bibitem[{Just} et~al.(2007){Just}, {Brandt}, {Shemmer} et~al.]{Just:2007cr}
{Just} D.~W., {Brandt} W.~N., {Shemmer} O., et~al., 2007, \apj, 665, 1004

\bibitem[{Kaspi} et~al.(2001){Kaspi}, {Brandt}, {Netzer} et~al.]{Kaspi:2001fu}
{Kaspi} S., {Brandt} W.~N., {Netzer} H., et~al., 2001, \apj, 554, 216

\bibitem[{Kelly} et~al.(2007){Kelly}, {Bechtold}, {Siemiginowska}, {Aldcroft}
  \& {Sobolewska}]{Kelly:2007fk}
{Kelly} B.~C., {Bechtold} J., {Siemiginowska} A., {Aldcroft} T., {Sobolewska}
  M., 2007, \apj, 657, 116

\bibitem[{Korista} et~al.(1998){Korista}, {Baldwin} \&
  {Ferland}]{Korista:1998dq}
{Korista} K., {Baldwin} J., {Ferland} G., 1998, \apj, 507, 24

\bibitem[{Krongold} et~al.(2009){Krongold}, {Jim{\'e}nez-Bail{\'o}n},
  {Santos-Lleo} et~al.]{Krongold:2009zr}
{Krongold} Y., {Jim{\'e}nez-Bail{\'o}n} E., {Santos-Lleo} M., et~al., 2009,
  \apj, 690, 773

\bibitem[{Krongold} et~al.(2005){Krongold}, {Nicastro}, {Elvis}, {Brickhouse},
  {Mathur} \& {Zezas}]{Krongold:2005jl}
{Krongold} Y., {Nicastro} F., {Elvis} M., {Brickhouse} N.~S., {Mathur} S.,
  {Zezas} A., 2005, \apj, 620, 165

\bibitem[{Larsson} et~al.(2008){Larsson}, {Fabian}, {Ballantyne} \&
  {Miniutti}]{Larsson:2008hc}
{Larsson} J., {Fabian} A.~C., {Ballantyne} D.~R., {Miniutti} G., 2008, \mnras,
  388, 1037

\bibitem[{Lee} et~al.(2002){Lee}, {Iwasawa}, {Houck}, {Fabian}, {Marshall} \&
  {Canizares}]{Lee:2002kl}
{Lee} J.~C., {Iwasawa} K., {Houck} J.~C., {Fabian} A.~C., {Marshall} H.~L.,
  {Canizares} C.~R., 2002, \apjl, 570, L47

\bibitem[{Liu} et~al.(2010){Liu}, {Elvis}, {McHardy} et~al.]{Liu:2010ys}
{Liu} Y., {Elvis} M., {McHardy} I.~M., et~al., 2010, \apj, 710, 1228

\bibitem[{Lobban} et~al.(2011){Lobban}, {Reeves}, {Miller}
  et~al.]{Lobban:2011fk}
{Lobban} A.~P., {Reeves} J.~N., {Miller} L., et~al., 2011, \mnras, 414, 1965

\bibitem[{Lobban} et~al.(2010){Lobban}, {Reeves}, {Porquet}
  et~al.]{Lobban:2010kl}
{Lobban} A.~P., {Reeves} J.~N., {Porquet} D., et~al., 2010, \mnras, 408, 551

\bibitem[{Longinotti} et~al.(2007){Longinotti}, {Bianchi}, {Santos-Lleo}
  et~al.]{Longinotti:2007ys}
{Longinotti} A.~L., {Bianchi} S., {Santos-Lleo} M., et~al., 2007, \aap, 470, 73

\bibitem[{Longinotti} et~al.(2003){Longinotti}, {Cappi}, {Nandra}, {Dadina} \&
  {Pellegrini}]{Longinotti:2003bs}
{Longinotti} A.~L., {Cappi} M., {Nandra} K., {Dadina} M., {Pellegrini} S.,
  2003, \aap, 410, 471

\bibitem[{Lubi{\'n}ski} \& {Zdziarski}(2001)]{Lubinski:2001kx}
{Lubi{\'n}ski} P., {Zdziarski} A.~A., 2001, \mnras, 323, L37

\bibitem[{Lubi{\'n}ski} et~al.(2010){Lubi{\'n}ski}, {Zdziarski}, {Walter}
  et~al.]{Lubinski:2010kx}
{Lubi{\'n}ski} P., {Zdziarski} A.~A., {Walter} R., et~al., 2010, \mnras, 408,
  1851

\bibitem[{Magdziarz} et~al.(1998){Magdziarz}, {Blaes}, {Zdziarski}, {Johnson}
  \& {Smith}]{Magdziarz:1998kl}
{Magdziarz} P., {Blaes} O.~M., {Zdziarski} A.~A., {Johnson} W.~N., {Smith}
  D.~A., 1998, \mnras, 301, 179

\bibitem[{Magdziarz} \& {Zdziarski}(1995)]{Magdziarz:1995fk}
{Magdziarz} P., {Zdziarski} A.~A., 1995, \mnras, 273, 837

\bibitem[{Marchese} et~al.(2012){Marchese}, {Della Ceca}, {Caccianiga},
  {Severgnini}, {Corral} \& {Fanali}]{Marchese:2012fk}
{Marchese} E., {Della Ceca} R., {Caccianiga} A., {Severgnini} P., {Corral} A.,
  {Fanali} R., 2012, \aap, 539, A48

\bibitem[{Marconi} et~al.(2004){Marconi}, {Risaliti}, {Gilli}, {Hunt},
  {Maiolino} \& {Salvati}]{Marconi:2004fk}
{Marconi} A., {Risaliti} G., {Gilli} R., {Hunt} L.~K., {Maiolino} R., {Salvati}
  M., 2004, \mnras, 351, 169

\bibitem[{Markowitz} et~al.(2009){Markowitz}, {Reeves}, {George}
  et~al.]{Markowitz:2009tg}
{Markowitz} A., {Reeves} J.~N., {George} I.~M., et~al., 2009, \apj, 691, 922

\bibitem[{Markowitz} et~al.(2008){Markowitz}, {Reeves}, {Miniutti}
  et~al.]{Markowitz:2008fv}
{Markowitz} A., {Reeves} J.~N., {Miniutti} G., et~al., 2008, \pasj, 60, 277

\bibitem[{Markowitz} \& {Reeves}(2009)]{Markowitz:2009nx}
{Markowitz} A.~G., {Reeves} J.~N., 2009, \apj, 705, 496

\bibitem[{Matsumoto} et~al.(2004){Matsumoto}, {Leighly} \&
  {Marshall}]{Matsumoto:2004pi}
{Matsumoto} C., {Leighly} K.~M., {Marshall} H.~L., 2004, \apj, 603, 456

\bibitem[{Mattson} et~al.(2007){Mattson}, {Weaver} \&
  {Reynolds}]{Mattson:2007bh}
{Mattson} B.~J., {Weaver} K.~A., {Reynolds} C.~S., 2007, \apj, 664, 101

\bibitem[{McKernan} \& {Yaqoob}(2004)]{McKernan:2004fv}
{McKernan} B., {Yaqoob} T., 2004, \apj, 608, 157

\bibitem[{McKernan} et~al.(2007){McKernan}, {Yaqoob} \&
  {Reynolds}]{McKernan:2007bh}
{McKernan} B., {Yaqoob} T., {Reynolds} C.~S., 2007, \mnras, 379, 1359

\bibitem[{Middleton} et~al.(2008){Middleton}, {Done} \&
  {Schurch}]{Middleton:2008ys}
{Middleton} M., {Done} C., {Schurch} N., 2008, \mnras, 383, 1501

\bibitem[{Miyazawa} et~al.(2009){Miyazawa}, {Haba} \&
  {Kunieda}]{Miyazawa:2009kl}
{Miyazawa} T., {Haba} Y., {Kunieda} H., 2009, \pasj, 61, 1331

\bibitem[{Mocz} et~al.(2011){Mocz}, {Lee}, {Iwasawa} \&
  {Canizares}]{Mocz:2011kl}
{Mocz} P., {Lee} J.~C., {Iwasawa} K., {Canizares} C.~R., 2011, \apj, 729, 30

\bibitem[{Murphy} \& {Yaqoob}(2009)]{Murphy:2009uq}
{Murphy} K.~D., {Yaqoob} T., 2009, \mnras, 397, 1549

\bibitem[{Nandra}(2006)]{Nandra:2006fk}
{Nandra} K., 2006, \mnras, 368, L62

\bibitem[{Nandra} et~al.(1997){Nandra}, {George}, {Mushotzky}, {Turner} \&
  {Yaqoob}]{Nandra:1997fk}
{Nandra} K., {George} I.~M., {Mushotzky} R.~F., {Turner} T.~J., {Yaqoob} T.,
  1997, \apjl, 488, L91

\bibitem[{Nandra} \& {Pounds}(1994)]{Nandra:1994ly}
{Nandra} K., {Pounds} K.~A., 1994, \mnras, 268, 405

\bibitem[{Nayakshin}(2000)]{Nayakshin:2000uq}
{Nayakshin} S., 2000, \apj, 534, 718

\bibitem[{Netzer} et~al.(2003){Netzer}, {Kaspi}, {Behar} et~al.]{Netzer:2003kl}
{Netzer} H., {Kaspi} S., {Behar} E., et~al., 2003, \apj, 599, 933

\bibitem[{Netzer} et~al.(2007){Netzer}, {Lira}, {Trakhtenbrot}, {Shemmer} \&
  {Cury}]{Netzer:2007vn}
{Netzer} H., {Lira} P., {Trakhtenbrot} B., {Shemmer} O., {Cury} I., 2007, \apj,
  671, 1256

\bibitem[{Noda} et~al.(2011){Noda}, {Makishima}, {Yamada}, {Torii}, {Sakurai}
  \& {Nakazawa}]{Noda:2011fv}
{Noda} H., {Makishima} K., {Yamada} S., {Torii} S., {Sakurai} S., {Nakazawa}
  K., 2011, \pasj, 63, 925

\bibitem[{Page} et~al.(2004){Page}, {O'Brien}, {Reeves} \&
  {Turner}]{Page:2004kx}
{Page} K.~L., {O'Brien} P.~T., {Reeves} J.~N., {Turner} M.~J.~L., 2004, \mnras,
  347, 316

\bibitem[{Page} et~al.(2005){Page}, {Reeves}, {O'Brien} \&
  {Turner}]{Page:2005qf}
{Page} K.~L., {Reeves} J.~N., {O'Brien} P.~T., {Turner} M.~J.~L., 2005, \mnras,
  364, 195

\bibitem[{Patrick} et~al.(2011){Patrick}, {Reeves}, {Porquet}, {Markowitz},
  {Lobban} \& {Terashima}]{Patrick:2011zt}
{Patrick} A.~R., {Reeves} J.~N., {Porquet} D., {Markowitz} A.~G., {Lobban}
  A.~P., {Terashima} Y., 2011, \mnras, 411, 2353

\bibitem[{Perola} et~al.(2002){Perola}, {Matt}, {Cappi} et~al.]{Perola:2002vn}
{Perola} G.~C., {Matt} G., {Cappi} M., et~al., 2002, \aap, 389, 802

\bibitem[{Peterson} et~al.(2004){Peterson}, {Ferrarese}, {Gilbert}
  et~al.]{Peterson:2004uq}
{Peterson} B.~M., {Ferrarese} L., {Gilbert} K.~M., et~al., 2004, \apj, 613, 682

\bibitem[{Petrucci} et~al.(2002){Petrucci}, {Henri}, {Maraschi}
  et~al.]{Petrucci:2002fk}
{Petrucci} P.~O., {Henri} G., {Maraschi} L., et~al., 2002, \aap, 388, L5

\bibitem[{Petrucci} et~al.(2013){Petrucci}, {Paltani}, {Malzac}
  et~al.]{Petrucci:2013fk}
{Petrucci} P.-O., {Paltani} S., {Malzac} J., et~al., 2013, \aap, 549, A73

\bibitem[{Ponti} et~al.(2009){Ponti}, {Cappi}, {Vignali} et~al.]{Ponti:2009bs}
{Ponti} G., {Cappi} M., {Vignali} C., et~al., 2009, \mnras, 394, 1487

\bibitem[{Ponti} et~al.(2006){Ponti}, {Miniutti}, {Cappi}, {Maraschi}, {Fabian}
  \& {Iwasawa}]{Ponti:2006uq}
{Ponti} G., {Miniutti} G., {Cappi} M., {Maraschi} L., {Fabian} A.~C., {Iwasawa}
  K., 2006, \mnras, 368, 903

\bibitem[{Porquet} et~al.(2004{\natexlab{a}}){Porquet}, {Kaastra}, {Page},
  {O'Brien}, {Ward} \& {Dubau}]{Porquet:2004ly}
{Porquet} D., {Kaastra} J.~S., {Page} K.~L., {O'Brien} P.~T., {Ward} M.~J.,
  {Dubau} J., 2004{\natexlab{a}}, \aap, 413, 913

\bibitem[{Porquet} et~al.(2004{\natexlab{b}}){Porquet}, {Reeves}, {O'Brien} \&
  {Brinkmann}]{Porquet:2004vn}
{Porquet} D., {Reeves} J.~N., {O'Brien} P., {Brinkmann} W., 2004{\natexlab{b}},
  \aap, 422, 85

\bibitem[{Press} et~al.(1992){Press}, {Teukolsky}, {Vetterling} \&
  {Flannery}]{Press:1992fk}
{Press} W.~H., {Teukolsky} S.~A., {Vetterling} W.~T., {Flannery} B.~P., 1992,
  {Numerical recipes in FORTRAN. The art of scientific computing}

\bibitem[{Qiao} \& {Liu}(2012)]{Qiao:2012uq}
{Qiao} E., {Liu} B.~F., 2012, \apj, 744, 145

\bibitem[{Ram{\'{\i}}rez} et~al.(2008){Ram{\'{\i}}rez}, {Komossa}, {Burwitz} \&
  {Mathur}]{Ramirez:2008fk}
{Ram{\'{\i}}rez} J.~M., {Komossa} S., {Burwitz} V., {Mathur} S., 2008, \apj,
  681, 965

\bibitem[{Reeves} et~al.(2007){Reeves}, {Awaki}, {Dewangan}
  et~al.]{Reeves:2007oq}
{Reeves} J.~N., {Awaki} H., {Dewangan} G.~C., et~al., 2007, \pasj, 59, 301

\bibitem[{Reeves} et~al.(2004){Reeves}, {Nandra}, {George}, {Pounds}, {Turner}
  \& {Yaqoob}]{Reeves:2004mi}
{Reeves} J.~N., {Nandra} K., {George} I.~M., {Pounds} K.~A., {Turner} T.~J.,
  {Yaqoob} T., 2004, \apj, 602, 648

\bibitem[{Reeves} et~al.(2009){Reeves}, {O'Brien}, {Braito}
  et~al.]{Reeves:2009qf}
{Reeves} J.~N., {O'Brien} P.~T., {Braito} V., et~al., 2009, \apj, 701, 493

\bibitem[{Reeves} et~al.(2003){Reeves}, {O'Brien} \& {Ward}]{Reeves:2003ve}
{Reeves} J.~N., {O'Brien} P.~T., {Ward} M.~J., 2003, \apjl, 593, L65

\bibitem[{Reeves} \& {Turner}(2000)]{Reeves:2000uq}
{Reeves} J.~N., {Turner} M.~J.~L., 2000, \mnras, 316, 234

\bibitem[{Reis} et~al.(2012){Reis}, {Fabian}, {Reynolds} et~al.]{Reis:2012pi}
{Reis} R.~C., {Fabian} A.~C., {Reynolds} C.~S., et~al., 2012, \apj, 745, 93

\bibitem[{Remillard} \& {McClintock}(2006)]{Remillard:2006uq}
{Remillard} R.~A., {McClintock} J.~E., 2006, \araa, 44, 49

\bibitem[{Ricci} et~al.(2013){Ricci}, {Paltani}, {Awaki}, {Petrucci}, {Ueda} \&
  {Brightman}]{Ricci:2013uq}
{Ricci} C., {Paltani} S., {Awaki} H., {Petrucci} P.-O., {Ueda} Y., {Brightman}
  M., 2013, \aap, 553, A29

\bibitem[{Ricci} et~al.(2011){Ricci}, {Walter}, {Courvoisier} \&
  {Paltani}]{Ricci:2011zr}
{Ricci} C., {Walter} R., {Courvoisier} T.~J.-L., {Paltani} S., 2011, \aap, 532,
  A102

\bibitem[{Risaliti} et~al.(2011){Risaliti}, {Nardini}, {Salvati}
  et~al.]{Risaliti:2011ly}
{Risaliti} G., {Nardini} E., {Salvati} M., et~al., 2011, \mnras, 410, 1027

\bibitem[{Risaliti} et~al.(2009){Risaliti}, {Young} \&
  {Elvis}]{Risaliti:2009nx}
{Risaliti} G., {Young} M., {Elvis} M., 2009, \apjl, 700, L6

\bibitem[{Russell} et~al.(2010){Russell}, {Fabian}, {Sanders}
  et~al.]{Russell:2010bh}
{Russell} H.~R., {Fabian} A.~C., {Sanders} J.~S., et~al., 2010, \mnras, 402,
  1561

\bibitem[{Saez} et~al.(2008){Saez}, {Chartas}, {Brandt} et~al.]{Saez:2008ys}
{Saez} C., {Chartas} G., {Brandt} W.~N., et~al., 2008, \aj, 135, 1505

\bibitem[{Sambruna} et~al.(2011){Sambruna}, {Tombesi}, {Reeves}
  et~al.]{Sambruna:2011oq}
{Sambruna} R.~M., {Tombesi} F., {Reeves} J.~N., et~al., 2011, \apj, 734, 105

\bibitem[{Schulze} \& {Wisotzki}(2010)]{Schulze:2010uq}
{Schulze} A., {Wisotzki} L., 2010, \aap, 516, A87

\bibitem[{Scott} et~al.(2005){Scott}, {Kriss}, {Lee} et~al.]{Scott:2005qo}
{Scott} J.~E., {Kriss} G.~A., {Lee} J.~C., et~al., 2005, \apj, 634, 193

\bibitem[{Shemmer} et~al.(2006){Shemmer}, {Brandt}, {Netzer}, {Maiolino} \&
  {Kaspi}]{Shemmer:2006uq}
{Shemmer} O., {Brandt} W.~N., {Netzer} H., {Maiolino} R., {Kaspi} S., 2006,
  \apjl, 646, L29

\bibitem[{Shemmer} et~al.(2008){Shemmer}, {Brandt}, {Netzer}, {Maiolino} \&
  {Kaspi}]{Shemmer:2008fk}
{Shemmer} O., {Brandt} W.~N., {Netzer} H., {Maiolino} R., {Kaspi} S., 2008,
  \apj, 682, 81

\bibitem[{Shemmer} et~al.(2005){Shemmer}, {Brandt}, {Vignali}
  et~al.]{Shemmer:2005dq}
{Shemmer} O., {Brandt} W.~N., {Vignali} C., et~al., 2005, \apj, 630, 729

\bibitem[{Shields}(2007)]{Shields:2007qf}
{Shields} J.~C., 2007, in { The Central Engine of Active Galactic Nuclei\/},
  edited by L.~C. {Ho}, J.-W. {Wang}, vol. 373 of { Astronomical Society of the
  Pacific Conference Series\/},  355

\bibitem[{Shu} et~al.(2012){Shu}, {Wang}, {Yaqoob}, {Jiang} \&
  {Zhou}]{Shu:2012fk}
{Shu} X.~W., {Wang} J.~X., {Yaqoob} T., {Jiang} P., {Zhou} Y.~Y., 2012, \apjl,
  744, L21

\bibitem[{Shu} et~al.(2010){Shu}, {Yaqoob} \& {Wang}]{Shu:2010zr}
{Shu} X.~W., {Yaqoob} T., {Wang} J.~X., 2010, \apjs, 187, 581

\bibitem[{Silva} et~al.(2004){Silva}, {Maiolino} \& {Granato}]{Silva:2004vn}
{Silva} L., {Maiolino} R., {Granato} G.~L., 2004, \mnras, 355, 973

\bibitem[{Singh} et~al.(2011){Singh}, {Shastri} \& {Risaliti}]{Singh:2011lh}
{Singh} V., {Shastri} P., {Risaliti} G., 2011, \aap, 532, A84

\bibitem[{Sobolewska} \& {Papadakis}(2009)]{Sobolewska:2009kx}
{Sobolewska} M.~A., {Papadakis} I.~E., 2009, \mnras, 399, 1597

\bibitem[{Soldi} et~al.(2008){Soldi}, {T{\"u}rler}, {Paltani}
  et~al.]{Soldi:2008qf}
{Soldi} S., {T{\"u}rler} M., {Paltani} S., et~al., 2008, \aap, 486, 411

\bibitem[{Starling} et~al.(2005){Starling}, {Page}, {Branduardi-Raymont},
  {Breeveld}, {Soria} \& {Wu}]{Starling:2005fu}
{Starling} R.~L.~C., {Page} M.~J., {Branduardi-Raymont} G., {Breeveld} A.~A.,
  {Soria} R., {Wu} K., 2005, \mnras, 356, 727

\bibitem[{Steenbrugge} et~al.(2003){Steenbrugge}, {Kaastra}, {Blustin}
  et~al.]{Steenbrugge:2003cr}
{Steenbrugge} K.~C., {Kaastra} J.~S., {Blustin} A.~J., et~al., 2003, \aap, 408,
  921

\bibitem[{Steenbrugge} et~al.(2005){Steenbrugge}, {Kaastra}, {Sako}
  et~al.]{Steenbrugge:2005fu}
{Steenbrugge} K.~C., {Kaastra} J.~S., {Sako} M., et~al., 2005, \aap, 432, 453

\bibitem[{Turner} et~al.(2005){Turner}, {Kraemer}, {George}, {Reeves} \&
  {Bottorff}]{Turner:2005ij}
{Turner} T.~J., {Kraemer} S.~B., {George} I.~M., {Reeves} J.~N., {Bottorff}
  M.~C., 2005, \apj, 618, 155

\bibitem[{Turner} et~al.(2011){Turner}, {Miller}, {Kraemer} \&
  {Reeves}]{Turner:2011bs}
{Turner} T.~J., {Miller} L., {Kraemer} S.~B., {Reeves} J.~N., 2011, \apj, 733,
  48

\bibitem[{Turner} et~al.(2007){Turner}, {Miller}, {Reeves} \&
  {Kraemer}]{Turner:2007ve}
{Turner} T.~J., {Miller} L., {Reeves} J.~N., {Kraemer} S.~B., 2007, \aap, 475,
  121

\bibitem[{Ueda} et~al.(2003){Ueda}, {Akiyama}, {Ohta} \& {Miyaji}]{Ueda:2003qf}
{Ueda} Y., {Akiyama} M., {Ohta} K., {Miyaji} T., 2003, \apj, 598, 886

\bibitem[{Vasudevan} \& {Fabian}(2007)]{Vasudevan:2007fk}
{Vasudevan} R.~V., {Fabian} A.~C., 2007, \mnras, 381, 1235

\bibitem[{Vasudevan} \& {Fabian}(2009)]{Vasudevan:2009vn}
{Vasudevan} R.~V., {Fabian} A.~C., 2009, \mnras, 392, 1124

\bibitem[{Vasudevan} et~al.(2010){Vasudevan}, {Fabian}, {Gandhi}, {Winter} \&
  {Mushotzky}]{Vasudevan:2010kx}
{Vasudevan} R.~V., {Fabian} A.~C., {Gandhi} P., {Winter} L.~M., {Mushotzky}
  R.~F., 2010, \mnras, 402, 1081

\bibitem[{Vasudevan} et~al.(2009){Vasudevan}, {Mushotzky}, {Winter} \&
  {Fabian}]{Vasudevan:2009uq}
{Vasudevan} R.~V., {Mushotzky} R.~F., {Winter} L.~M., {Fabian} A.~C., 2009,
  \mnras, 399, 1553

\bibitem[{Vestergaard}(2002)]{Vestergaard:2002vn}
{Vestergaard} M., 2002, \apj, 571, 733

\bibitem[{Vignali} et~al.(2004){Vignali}, {Brandt}, {Boller}, {Fabian} \&
  {Vaughan}]{Vignali:2004ff}
{Vignali} C., {Brandt} W.~N., {Boller} T., {Fabian} A.~C., {Vaughan} S., 2004,
  \mnras, 347, 854

\bibitem[{Vignali} et~al.(2005){Vignali}, {Brandt}, {Schneider} \&
  {Kaspi}]{Vignali:2005bh}
{Vignali} C., {Brandt} W.~N., {Schneider} D.~P., {Kaspi} S., 2005, \aj, 129,
  2519

\bibitem[{Wang} et~al.(2009){Wang}, {Mao} \& {Wei}]{Wang:2009ly}
{Wang} J., {Mao} Y.~F., {Wei} J.~Y., 2009, \aj, 137, 3388

\bibitem[{Wang} et~al.(2010){Wang}, {Risaliti}, {Fabbiano}, {Elvis}, {Zezas} \&
  {Karovska}]{Wang:2010vn}
{Wang} J., {Risaliti} G., {Fabbiano} G., {Elvis} M., {Zezas} A., {Karovska} M.,
  2010, \apj, 714, 1497

\bibitem[{Wang} et~al.(2004){Wang}, {Watarai} \& {Mineshige}]{Wang:2004ly}
{Wang} J.-M., {Watarai} K.-Y., {Mineshige} S., 2004, \apjl, 607, L107

\bibitem[{Wilms} et~al.(2001){Wilms}, {Reynolds}, {Begelman}
  et~al.]{Wilms:2001tg}
{Wilms} J., {Reynolds} C.~S., {Begelman} M.~C., et~al., 2001, \mnras, 328, L27

\bibitem[{Winter} et~al.(2009){Winter}, {Mushotzky}, {Reynolds} \&
  {Tueller}]{Winter:2009uq}
{Winter} L.~M., {Mushotzky} R.~F., {Reynolds} C.~S., {Tueller} J., 2009, \apj,
  690, 1322

\bibitem[{Winter} et~al.(2012){Winter}, {Veilleux}, {McKernan} \&
  {Kallman}]{Winter:2012ve}
{Winter} L.~M., {Veilleux} S., {McKernan} B., {Kallman} T.~R., 2012, \apj, 745,
  107

\bibitem[{Woo} \& {Urry}(2002)]{Woo:2002ve}
{Woo} J.-H., {Urry} C.~M., 2002, \apj, 579, 530

\bibitem[{Yamaoka} et~al.(2005){Yamaoka}, {Uzawa}, {Arai}, {Yamazaki} \&
  {Yoshida}]{Yamaoka:2005ys}
{Yamaoka} K., {Uzawa} M., {Arai} M., {Yamazaki} T., {Yoshida} A., 2005, Chinese
  Journal of Astronomy and Astrophysics Supplement, 5, 273

\bibitem[{Yaqoob} et~al.(2003{\natexlab{a}}){Yaqoob}, {George}, {Kallman},
  {Padmanabhan}, {Weaver} \& {Turner}]{Yaqoob:2003qa}
{Yaqoob} T., {George} I.~M., {Kallman} T.~R., {Padmanabhan} U., {Weaver} K.~A.,
  {Turner} T.~J., 2003{\natexlab{a}}, \apj, 596, 85

\bibitem[{Yaqoob} et~al.(2003{\natexlab{b}}){Yaqoob}, {McKernan}, {Kraemer}
  et~al.]{Yaqoob:2003ij}
{Yaqoob} T., {McKernan} B., {Kraemer} S.~B., et~al., 2003{\natexlab{b}}, \apj,
  582, 105

\bibitem[{Younes} et~al.(2011){Younes}, {Porquet}, {Sabra} \&
  {Reeves}]{Younes:2011cr}
{Younes} G., {Porquet} D., {Sabra} B., {Reeves} J.~N., 2011, \aap, 530, A149

\bibitem[{Young} et~al.(2009){Young}, {Elvis} \& {Risaliti}]{Young:2009uq}
{Young} M., {Elvis} M., {Risaliti} G., 2009, \apjs, 183, 17

\bibitem[{Zdziarski} et~al.(2003){Zdziarski}, {Lubi{\'n}ski}, {Gilfanov} \&
  {Revnivtsev}]{Zdziarski:2003tg}
{Zdziarski} A.~A., {Lubi{\'n}ski} P., {Gilfanov} M., {Revnivtsev} M., 2003,
  \mnras, 342, 355

\bibitem[{Zhang} et~al.(2011){Zhang}, {Ji}, {Marshall}, {Longinotti}, {Evans}
  \& {Gu}]{Zhang:2011zr}
{Zhang} S.~N., {Ji} L., {Marshall} H.~L., {Longinotti} A.~L., {Evans} D., {Gu}
  Q.~S., 2011, \mnras, 410, 2274

\end{thebibliography}
 
 \end{document}